\pdfoutput=1
\documentclass[aps,superscriptaddress,tightenlines,nofootinbib,floatfix,longbibliography,notitlepage]{revtex4-1}
\usepackage[left=18mm,right=19mm,top=23mm,bottom=16mm]{geometry}

\providecommand{\repositoryInformationSetup}{} %
\repositoryInformationSetup

\usepackage{xspace}
\usepackage{bbm}
\usepackage{amsmath,amssymb}
\usepackage{mathtools}
\usepackage{dsfont}
\usepackage{braket}
\usepackage[T1]{fontenc}
\usepackage[boxruled,lined,commentsnumbered]{algorithm2e}
\usepackage[utf8]{inputenc}
\usepackage[UKenglish]{babel}
\usepackage{siunitx}
\usepackage{placeins}
\usepackage{bm}
\usepackage{graphicx}
\usepackage[dvipsnames]{xcolor}
\usepackage{slashed}
\usepackage{xfrac}
\usepackage{changes}
\usepackage[
colorlinks=true,
allcolors=blue
]{hyperref}

\graphicspath{{data/},{figures/}}

\newcommand{\repoURL}{https://github.com/evanberkowitz/hubbard-critical-coupling}

\newcommand{\Secref}[1]{Section~\ref{sec:#1}}

\newcommand{\Appref}[1]{Appendix~\ref{sec:#1}}

\newcommand{\Figref}[1]{Figure~\ref{fig:#1}\xspace}

\renewcommand{\Ref}[1]{Ref.~\cite{#1}}

\newcommand{\Refs}[1]{Refs.~\cite{#1}}

\newcommand{\goesto}{\ensuremath{\rightarrow}}

\DeclareMathOperator{\tr}{Tr}

\newcommand{\inverse}{\ensuremath{^{-1}}}                                       %
\DeclareMathOperator{\ord}{\mathcal{O}}

\newcommand{\ordnung}[1]{\ensuremath{\ord(#1)}}
\newcommand{\efford}[1]{\ensuremath{\ord_{\text{e}}(#1)}}
\newcommand{\erwartung}[1]{\ensuremath{\left\langle#1\right\rangle}}

\newcommand{\pdagger}{{\phantom{\dagger}}}
\newcommand{\kappadelta}{\ensuremath{\tilde{\kappa}}}
\newcommand{\phidelta}{\ensuremath{\tilde{\phi}}}
\newcommand{\mdelta}{\ensuremath{\tilde{m}_s}}

\newcommand{\meff}{\ensuremath{m_\text{eff}}\xspace}
\newcommand{\Z}{\ensuremath{\mathcal{Z}}}
\newcommand{\D}{\ensuremath{\mathcal{D} }}
\newcommand{\nt}{\ensuremath{N_t}\xspace}

\newcommand{\up}{\ensuremath{\uparrow}}
\newcommand{\down}{\ensuremath{\downarrow}}

\newcommand{\matr}[2]{\left(\begin{array}{#1}#2\end{array}\right)}

\newcommand{\delsqu}[2]{\ensuremath{\frac{\partial^2 #1}{\partial #2 ^2}}}

\newcommand{\adjoint}{\ensuremath{{}^{\dagger}}}
\newcommand{\conjugate}{\ensuremath{{}^{*}}}

\newcommand{\effm}{effective mass}

\newcommand{\spc}{single-particle correlator}

\let\builtinLaTeX\LaTeX
\def\LaTeX{\builtinLaTeX\xspace}

   \newcommand{\critU}{\ensuremath{\num{3.834(14)}}}
  \newcommand{\critMu}{\ensuremath{\num{0.844(31)}}}
  \newcommand{\critNu}{\ensuremath{\num{1.185(43)}}} %
\newcommand{\critZeta}{\ensuremath{\num{0.924(17)}}}
 \newcommand{\critExp}{\ensuremath{\num{1.095(37)}}}

  \newcommand{\critUf}{\ensuremath{\num{3.877(46)}}}
  \newcommand{\critUg}{\ensuremath{\num{3.876(79)}}}

\newcommand{\DeltaInf}{\ensuremath{\num{0.057(11)}}}
  \newcommand{\coeffU}{\ensuremath{\num{0.479(93)}}}

\newcommand{\bonn}{
    Helmholtz-Institut f\"{u}r Strahlen- und Kernphysik,
    Rheinische Friedrich-Wilhelms-Universit\"{a}t Bonn, 53012 Bonn, Germany
}

\newcommand{\ikp}{
    Institut f\"{u}r Kernphysik,
    Forschungszentrum J\"{u}lich, 54245 J\"{u}lich, Germany
}

\newcommand{\ias}{
    Institute for Advanced Simulation,
    Forschungszentrum J\"{u}lich, 54245 J\"{u}lich, Germany
}

\newcommand{\jsc}{
    J\"{u}lich Supercomputing Center,
    Forschungszentrum J\"{u}lich, 54245 J\"{u}lich, Germany
}

\newcommand{\mcfp}{
    Maryland Center for Fundamental Physics,
    University of Maryland, College Park 20742, USA
}
 
\begin{document}

\title{The Semimetal-Mott Insulator Quantum Phase Transition of the Hubbard Model on the Honeycomb Lattice}

\author{Johann Ostmeyer}        \affiliation{\bonn}
\author{Evan Berkowitz}         \affiliation{\mcfp}\affiliation{\ias}
\author{Stefan Krieg}           \affiliation{\jsc} \affiliation{\ias}
\author{Timo A. L\"{a}hde}      \affiliation{\ias} \affiliation{\ikp}
\author{Thomas Luu}             \affiliation{\ias} \affiliation{\ikp} \affiliation{\bonn}
\author{Carsten Urbach}         \affiliation{\bonn}

\date{\today}

\begin{abstract}
We take advantage of recent improvements in the grand canonical Hybrid Monte Carlo algorithm, to perform
a precision study of the single-particle gap in the hexagonal Hubbard model, with on-site electron-electron interactions. After
carefully controlled analyses of the Trotter error, the thermodynamic limit, and finite-size scaling with inverse temperature, we find a
critical coupling of $U_c/\kappa=\critU$ and the critical exponent $z\nu=\critNu$. Under the assumption that this corresponds to the expected anti-ferromagnetic Mott transition, we
are also able to provide a preliminary estimate $\beta=\critExp$ for the critical exponent of the order parameter. We consider our findings in view
of the $SU(2)$ Gross-Neveu, or chiral Heisenberg, universality class. We also discuss the computational scaling of the Hybrid Monte Carlo 
algorithm, and possible extensions of our work to carbon nanotubes, fullerenes, and topological insulators.
\end{abstract}

\maketitle

\allowdisplaybreaks[1]

\section{Introduction 
\label{sec:intro}}

Monte Carlo (MC) simulations of strongly correlated electrons in carbon nano-materials~\cite{GeimNovoselovReview,CastroNeto2009,Kotov2012} is an 
emerging topic in both the condensed matter~\cite{Khveshchenko2004323,heisenberg_gross_neveu,PhysRevB.86.115447} and
nuclear physics communities~\cite{Drut:2008rg,Hands:2008id}. The basis of such studies is the Hubbard model, 
a Hamiltonian approach which reduces, at weak electron-electron coupling, to the tight-binding description of atomic orbitals in a lattice 
of carbon ions~\cite{Saito1998,Wehling2011,Tang2015}. The properties of the Hubbard model on a honeycomb lattice are thought to resemble those of graphene.
MC simulations of the Hubbard model are closely related to 
problems of current interest in atomic and nuclear physics, such as the unitary Fermi gas~\cite{Chen:2003vy,Bulgac:2005pj,Bloch:2008zzb,Drut:2010yn}
and nuclear lattice effective field theory~\cite{Borasoy:2006qn,Lee:2008fa,Lahde:2013uqa,Lahde:2019npb}.

Our objective is to take advantage of this recent algorithmic development and perform, for the first time, a precision calculation of the single-particle
gap $\Delta$ of the hexagonal Hubbard model, in the grand canonical ensemble. Whether such a gap exists or not, is determined by
the relative strength of the on-site electron-electron coupling $U$, and the nearest-neighbor hopping amplitude $\kappa$.
Prior MC work in the canonical ensemble has established the existence of a second-order transition into a gapped, anti-ferromagnetic, Mott insulating (AFMI) phase
at a critical coupling of $U_c/\kappa \simeq 3.8$~\cite{Meng2010,Assaad:2013xua,Wang:2014cbw,Otsuka:2015iba}. The existence of an intermediate spin-liquid (SL)
phase~\cite{Meng2010} at couplings slightly below $U_c$, appears to now be disfavored~\cite{Otsuka:2015iba}.
Long-range interactions in graphene~\cite{Son:2007ja,Smith:2014tha} are thought to frustrate the AFMI transition, as these
favor charge-density wave (CDW) symmetry breaking~\cite{Buividovich:2016tgo,Buividovich:2018yar} over AFMI. The critical exponents of the AFMI transition should 
fall into the $SU(2)$ Gross-Neveu (GN), or chiral Heisenberg, universality class~\cite{gross_neveu_orig,heisenberg_gross_neveu}.

The observability of the AFMI transition in graphene is of interest for fundamental as well as applied physics.
While $\kappa$ is well constrained from density functional theory (DFT) and experiment~\cite{CastroNeto2009}, the on-site coupling $U$ is more difficult to 
determine theoretically for graphene~\cite{Tang2015}, although the physical value of $U/\kappa$ is commonly believed to be insufficient to trigger the AFMI 
phase in samples of suspended graphene or with application of biaxial strain~\cite{Wehling2011,Tang2015}. The AFMI phase may be more easily observed in the presence
of external magnetic fields~\cite{Gorbar:2002iw,Herbut:2008ui}, and the Fermi velocity at the Dirac point may still be strongly renormalized due to interaction effects~\cite{Tang2015}.
The reduced dimensionality of fullerenes and carbon nanotubes may increase the importance of electron-electron interactions in such systems~\cite{Luu:2015gpl}.

Let us summarize the layout of our paper. Our lattice fermion operator and Hybrid Monte Carlo (HMC) algorithm is introduced in \Secref{formalism}.
We describe in \Secref{results} how correlation functions and $\Delta$ are computed from MC simulations in the grand canonical ensemble. We also give details on our
extrapolation in the temporal lattice spacing (or Trotter error) $\delta$ and system (lattice) size $L$, and provide results for $\Delta$ as a function of $U/\kappa$ and inverse temperature $\beta$.
In \Secref{analysis}, we analyze these results using finite-size scaling in $\beta$, and provide our best estimate $U_c/\kappa=\critU$ for the critical coupling at which $\Delta$ becomes
non-zero. We also provide a preliminary estimate of the critical exponent of the AFMI order parameter, under the assumption that
the opening of the gap coincides with the AFMI transition. In \Secref{conclusion}, we compare our results with other studies of the Hubbard model and the chiral Heisenberg universality class, 
and discuss possible extensions of our work to carbon nanotubes, fullerenes and topological insulators.

\section{Formalism 
\label{sec:formalism}}

The Hubbard model is a theory of interacting fermions, that can hop between nearest-neighbor sites. We focus on the two-dimensional honeycomb lattice, 
which is bipartite in terms of $A$~sites and $B$~sites. The Hamiltonian is given by
\begin{equation}
H := - \sum_{xy} 
(a\adjoint_x h_{xy}^{} a_y^{} + b\adjoint_x h_{xy}^{} b_y^{})
+ \frac{1}{2} \sum_{xy} \rho_x^{} V_{xy}^{} \rho_y^{},
\label{Ham_def}
\end{equation}
where we have applied a particle-hole transformation to a theory of spin \up\ and spin \down\ electrons. Here, 
$a^\dag$ and $a$ are creation and annihilation operators for particles (spin-up electrons), and $b^\dag$ and $b$ are similarly
for holes (spin-down electrons). As usual, the signs of the $b$ operators have been switched for $B$~sites.
The matrix $h_{xy} := \kappa \delta_{\langle x,y \rangle}$ describes nearest-neighbor hopping, while $V_{xy}$ is the potential between particles on different sites, 
and $\rho_x := b\adjoint_x b_x - a\adjoint_x a_x$ is the charge operator. In this work, we study the Hubbard model with on-site interactions only, such that $V_{xy} = U\delta_{xy}$;
the ratio $U/\kappa$ determines whether we are in a strongly or weakly coupled regime.

Hamiltonian theories such as~(\ref{Ham_def}) have for a long time been studied with lattice MC methods~\cite{Paiva2005,Beyl:2017kwp}, as this allows for 
a fully \textit{ab initio} stochastic evaluation of the thermal trace, or Grassmann path integral. There is a large freedom of choice in the construction of lattice MC algorithms, 
including the discretization of the theory, the choice of Hubbard-Stratonovich (or auxiliary field) transformation, and the algorithm used to update the auxiliary field variables.
This freedom can be exploited to optimize the algorithm with respect to a particular computational feature. These pertain to the scaling of the computational effort with system (lattice) 
size $L$, inverse temperature $\beta$, number of time slices $N_t=\beta/\delta$, interaction strength $U/\kappa$, and electron number density (for simulations away from half filling).
Hamiltonian theories are often simulated with an exponential (or compact) form of both the kinetic and 
potential energy contributions to the partition function (or Euclidean time projection amplitude), and with random Metropolis updates of the auxiliary fields (which may be either 
discrete or continuous). In condensed matter and atomic physics, such methods are referred to as the Blankenbecler-Sugar-Scalapino (BSS) 
algorithm~\cite{Blankenbecler:1981jt}.

In Lattice Quantum Chromodynamics (QCD), the high dimensionality of the theory and the need to precisely approach the continuum limit have led to the development of specialized algorithms 
which optimize the computational scaling with $L$. These efforts have culminated in the HMC algorithm, which combines elements of the 
Langevin, Molecular Dynamics (MD), and Metropolis algorithms~\cite{Duane:1987de}. The application of HMC to the Hubbard model~(\ref{Ham_def}) has proven to be 
surprisingly difficult, due to problems related to ergodicity, symmetries of the Hamiltonian, and the correct approach to the (temporal) continuum limit.
For a thorough treatment of these from the point of view of HMC, see \Refs{Luu:2015gpl,Wynen:2018ryx,Ostmeyer:thesis}.
In order to realize the expected $\sim V^{5/4}$ computational scaling (where $V = 2L^2$), a suitable conjugate gradient (CG) 
method has to be found for the numerical integration of the MD equations of motion.
The Hasenbusch preconditioner~\cite{hasenbusch} from Lattice QCD has recently been found to work for the Hubbard model as well~\cite{2018arXiv180407195K}.
The resulting combination of HMC with the Hubbard model is referred to as the Brower-Rebbi-Schaich (BRS) algorithm~\cite{Brower:2011av,Brower:2012zd}, which is closely related to the 
BSS algorithm. The main differences are the linearized kinetic energy (or nearest-neighbor hopping) term, and the purely imaginary auxiliary field, which is updated using HMC moves in the
BRS algorithm.

The BSS algorithm has preferentially been used within the canonical ensemble~\cite{Meng2010,Assaad:2013xua,Otsuka:2015iba}. This entails a
projection Monte Carlo (PMC) calculation, where particle number is conserved and one is restricted to specific many-body Hilbert spaces.
PMC is highly efficient at accessing zero-temperature (or ground state) properties, especially when the number of particles is constant, for
instance the $A$ nucleons in an atomic nucleus. For the Hubbard model on the honeycomb lattice at half-filling, the fully anti-symmetric trial wave function 
encodes a basis of $2L^2$ electrons to be propagated in Euclidean time. Due to this scaling of the number of trial wave functions, PMC and grand canonical 
versions of the BSS algorithm both exhibit $\sim V^3$ scaling (with random, local Metropolis updates). In contrast to PMC, the grand canonical
formalism resides in the full Fock space, and no trial wave function is used. Instead, Boltzmann-weighted thermal expectation values of observables are extracted.
At low temperatures and large Euclidean times, spectral observables are measured relative to the ground state of the Fock space, which is the half-filling state 
(an explicit example is given in \Secref{results}).
With HMC updates, such an algorithm has been found to scale as $\sim V^{5/4}$~\cite{Creutz:1988wv,2018arXiv180407195K}.
A drawback of the grand canonical ensemble is the explicit inverse temperature $\beta$. Thus, the limit $\beta \to \infty$ is taken by extrapolation, or
more specifically by finite-size scaling. This $\beta$-dependence may be considerable, though observable-dependent. While PMC simulations are 
not free of similar effects (due to contamination from excited state contributions), 
they are typically less severe due to the absence of backwards-propagating states in Euclidean time.

For a number of reasons, HMC updates have proven difficult for the BSS algorithm. First, the exponential form of the fermion operator $M$ causes $\det M$
to factorize into regions of positive and negative sign. Though this does not imply a sign problem at half filling (the action $S \sim |\det M|^2$), 
it does introduce boundaries in the energy landscape of the theory, which HMC trajectories in general cannot cross without special and computationally very expensive 
methods~\cite{Fodor:2003bh,Cundy:2005pi}. For $\beta \to \infty$, this fragmentation effect increases dramatically, and causes an ergodicity problem with HMC.
Second, while this problem can be circumvented by a complex-valued auxiliary field, the resulting ``complexified'' HMC algorithm shows poor (roughly cubic) scaling with $V$~\cite{Beyl:2017kwp}.
It is interesting to note how the BRS algorithm avoids this ergodicity problem. Due to the linearization of the hopping term in the fermion operator (with imaginary auxiliary field), 
the boundaries impassable to HMC are reduced in dimension and can be avoided~\cite{Wynen:2018ryx}. Naturally, the BSS and BRS formulations become equivalent in the temporal
continuum limit. Then, the ergodicity problem would eventually be recovered when $\delta \to 0$ (where MC simulations are in any case not practical).
A drawback specific to BRS is that spin symmetry is explicitly broken for $\delta \neq 0$, due to the linearization of the hopping term~\cite{Buividovich:2016tgo}.
Hence, the choice of BSS versus BRS represents a tradeoff between the retention of more symmetries at finite $\delta$, and faster convergence to the 
continuum limit (BSS), or improved ergodicity and computational scaling with $V$ (BRS).

Here, we apply the BRS algorithm with HMC updates to the Hubbard model~(\ref{Ham_def}). This entails the stochastic evaluation of a path integral over a Hubbard-Stratonovich
field $\phi$.
The exact form of the fermion operator $M$ depends on the choice of discretization
for time derivatives. We adopt the conventions of \Ref{Luu:2015gpl}, which used a ``mixed differencing'' operator, with forward differencing in time for $A$ sites, and backward 
for $B$ sites. For reference, we note that this scheme does not introduce a fermion doubling problem. For the non-interacting theory, mixed differencing gives $\ordnung{\delta^2}$ discretization 
errors (per time step). Numerically, mixed differencing has been shown to approach the limit $\delta \to 0$ faster than pure forward or backward differencing when $U > 0$. 
The explicit form of $M$ is
\begin{equation}
\begin{split}
M^{AA}_{(x,t)(y,t')} & = \delta_{xy}^{} \left(\delta_{t+1,t^\prime}^{} - \delta_{t,t^\prime}^{} \exp(-i\phidelta_{x,t}^{}) \right), \\
M^{BB}_{(x,t)(y,t')} & = \delta_{xy}^{} \left(\delta_{t,t^\prime} - \delta_{t-1,t'} \exp(-i\phidelta_{x,t}^{}) \right), \\
M^{AB}_{(x,t)(y,t')} = M^{BA}_{(x,t)(y,t^\prime)} & =- \kappadelta\,\delta_{\erwartung{x,y}} \delta_{t,t^\prime},
\end{split}
\label{eqn_ferm_op}
\end{equation}
where the dependence on $A$ and $B$ sites has been written out. All quantities multiplied by $\delta$ have been denoted by a tilde. While the hopping term
in Eq.~(\ref{eqn_ferm_op}) has been linearized, the auxiliary field $\tilde\phi$ enters through exponential ``gauge links'' familiar from Lattice QCD. As first discussed 
in Refs.~\cite{Brower:2011av,Brower:2012zd}, such gauge links contain a ``seagull term'', which needs to be correctly handled in order to recover the physical $\delta \to 0$ limit.
This condition is satisfied by Eq.~(\ref{eqn_ferm_op}), and further details are given in \Appref{mixed_bias}. 

\section{The gap 
\label{sec:results}}

\subsection{The \spc}

We shall now describe the procedure of obtaining the single-particle gap $\Delta$ as a function of $U/\kappa$ and $\beta$, from a given ensemble of auxiliary-field 
configurations. We recall that $a$ and $a\adjoint$ are annihilation and creation operators for quasiparticles, and similarly $b$ and $b\adjoint$ for (quasi-)holes.
For instance, by creating and destroying quasiparticles at different locations and times, we obtain the correlator
\begin{equation}
C_{xy}^{}(t)
:= \left\langle a_{x,t}^\pdagger a\adjoint_{y,0} \right\rangle
= \frac{1}{\Z} \int \D\phi\; M[\phi] \inverse_{(x,t),(y,0)} \exp(-S[\phi])
= \left\langle M[\phi] \inverse_{(x,t),(y,0)} \right\rangle,
\end{equation}
as an ensemble average, 
where $S[\phi]$ is the Euclidean action and $\Z$ is the partition function---the integral without $M$.
We have used Wick contraction to replace the operators with the fermion propagator.

We now move to the Heisenberg picture and express the correlator 
as a thermal trace
\begin{equation}
C_{xy}(t) = \left\langle a_{x,t}^\pdagger a\adjoint_{y,0} \right\rangle
= \frac{1}{\Z} \tr\bigg\{a_{x,t}^\pdagger a\adjoint_{y,0} \exp(-\beta H) \bigg\}
= \frac{1}{\Z} \tr\bigg\{\exp(-H(\beta-t)) \, a_x^\pdagger \exp(-Ht) a\adjoint_y \bigg\},
\end{equation}
and by inserting the identity (resolved in the interacting basis), we find the spectral decomposition
\begin{align}
\label{eq:spectral decomposition}
C_{xy}^{}(t) = \left\langle a_{x,t}^\pdagger a\adjoint_{y,0} \right\rangle & =
\frac{1}{\sum_i \exp(-\beta E_i)} 
\sum_{m,n} \exp(-\beta E_m) \exp(-(E_n-E_m)t) 
\, z_{mxn}^{\vphantom{*}} z\conjugate_{myn}, \\
\label{eq:overlap factor}
z_{mxn}^{} & :=
\left\langle m \middle| a_x^{} \middle| n \right\rangle,
\end{align}
where the summation indices $m$ and $n$ denote interacting eigenstates with energies $E_m^{}$ and $E_n^{}$, respectively, and the $z_{ijk}^{}$ are referred to as ``overlap factors''.
At large $\beta$ and in the limit of large Euclidean time, the correlator decays exponentially,
\begin{equation}
\lim_{t\goesto\infty} \lim_{\beta\goesto\infty} 
C_{xy}^{}(t) \simeq \exp(-(E_1-E_0)t) \, z_{0x1}^{\vphantom{*}} z\conjugate_{0y1},
\end{equation}
where the energy $E_1^{}$ is measured relative to the energy $E_0$ of ground state of the Fock space, assuming that the associated 
overlap factors are non-zero. 

To continue it is convenient to Fourier transform the fermion propagator, which is a function of the lattice coordinates $x$ and $y$, to a function of the (lattice) momenta $k$ and $p$\footnote{We note that under the ensemble average the correlators are diagonal in $k$ and $p$ due to translational invariance.},
\begin{equation}
C_{kp}^{}(t)=\frac{1}{L^4}\sum_{x,y}\exp(-ikx-ipy) \, C_{xy}^{}(t)\ .
\end{equation}
Repetition of this procedure yields an ensemble of ``measurements'', the average of which is our MC estimate of $C_{kp}(t)$.
As there are two Dirac points $K$ and $K^\prime$, by symmetry we have\footnote{Because of the underlying sublattices $A$ and $B$, 
there are in principle two independent correlators for each momentum $k$~\cite{Luu:2015gpl}.  However, these two correlators 
are degenerate at each Dirac point $K$ and $K^\prime$; we average them to construct $C_{KK}$ and $C_{K^\prime K^\prime}$, respectively.}
\begin{equation}
C_{KK}^{}(t) = C_{K^\prime K^\prime}^{}(t),
\end{equation}
which holds for the expectation values, but not on a configuration-by-configuration basis. By choosing to Fourier transform $x$ and $y$ to the Dirac momenta $K$ or $K^\prime$, we adjust the overlap factors so that we can take
$E_1^{}$ to refer to the energy at the Dirac point. We define the ``effective mass''
\begin{equation}
\meff := E_1^{}-E_0^{},
\end{equation}
which can be extracted from the correlator according to
\begin{equation}
\label{eq:usual effective mass}
\meff(t) = -\lim_{\beta\goesto\infty} \partial_t \ln C_{KK}^{}(t),
\qquad
\meff = \lim_{t\goesto\infty} \meff(t),
\end{equation}
such that the single-particle gap is given by
\begin{equation}
\Delta = 2 \meff,
\end{equation}
for a specific value of $L$ and $\delta$. We discuss the extrapolations in these variables in Section~\ref{sec:gap_results}.

\begin{figure}[t]
\centering
\includegraphics[width=.45\textwidth]{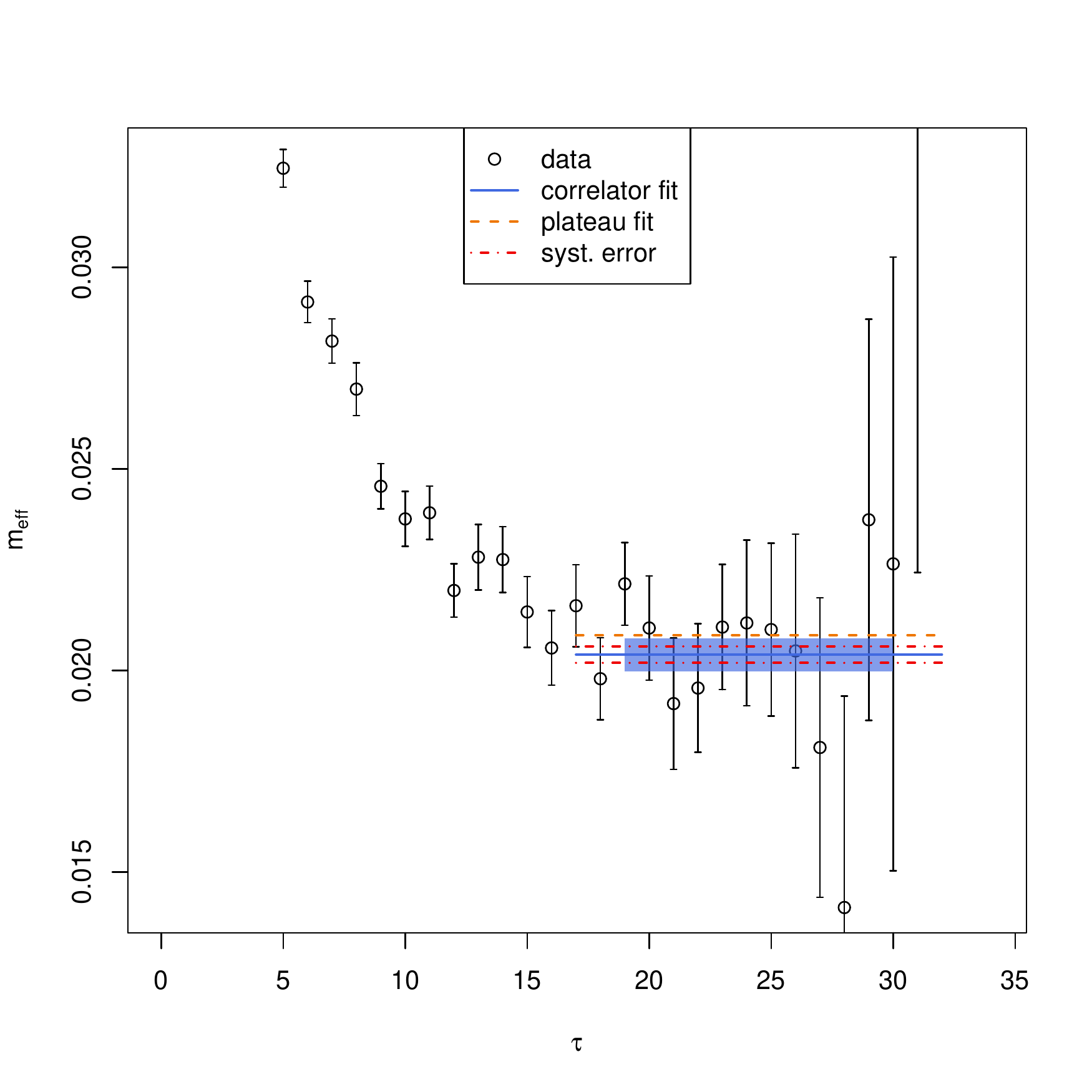}
\includegraphics[width=.45\textwidth]{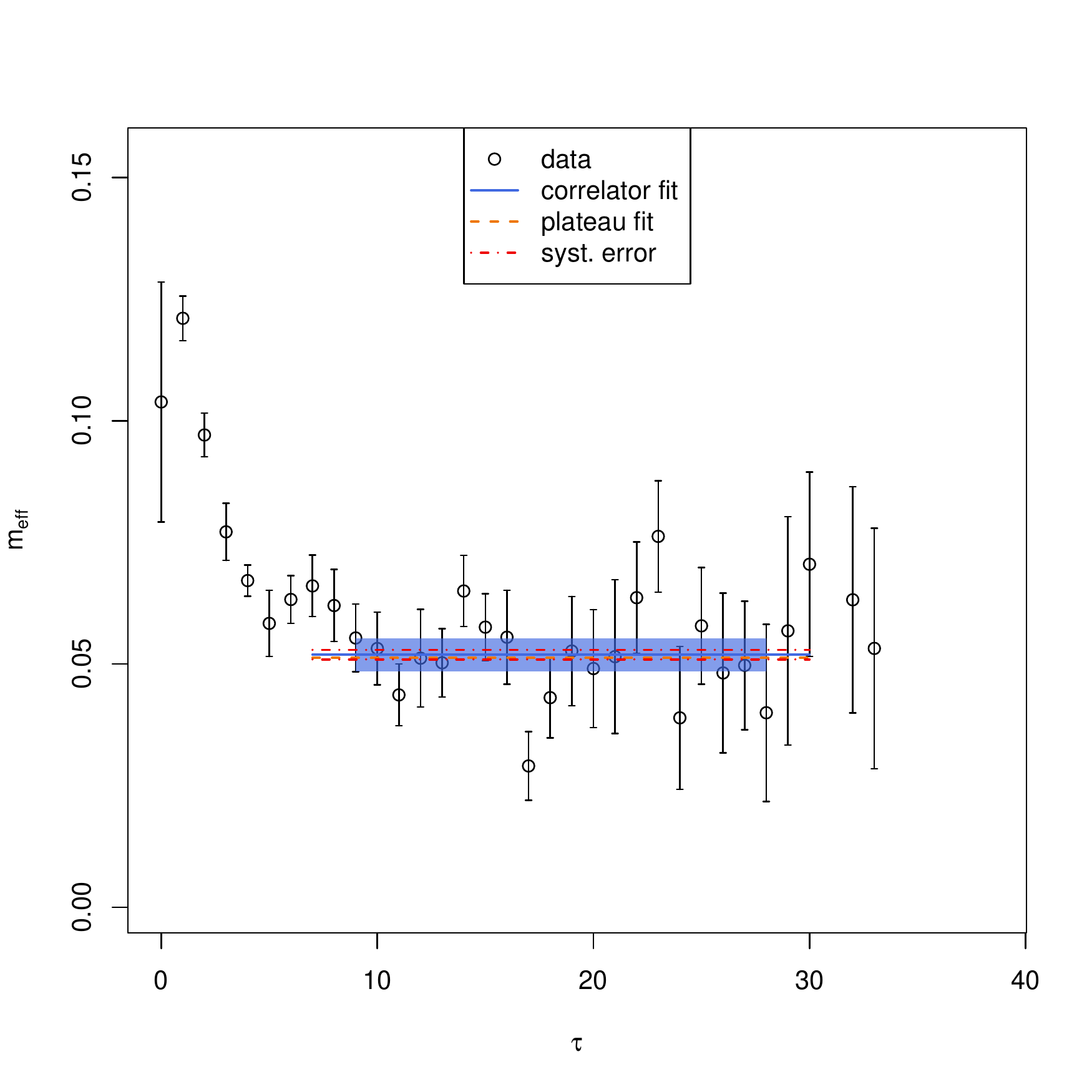}
\caption{Examples of effective mass determinations from the {\spc}s, extracted from ensembles of auxiliary field configurations.
The blue line with error band gives $\meff$ with statistical error, obtained from a hyperbolic cosine effective mass \eqref{eq:cosh fit}.
The length of the blue band indicates the fitting region. For comparison, a constant fit to the effective mass \eqref{eq:usual effective mass} is 
shown by the dashed orange line. The dot-dashed red line shows the estimation of the systematic error, as explained in \Appref{plateau_estimation}.
Note that the red and orange lines have been extended outside of the fitting region, for clearer visibility.
Left panel: $\kappa\beta=8$, $L=15$, $N_t=64$, $U/\kappa=\num{3.5}$. Right panel: $\kappa\beta=12$, $L=6$, $N_t=72$, $U/\kappa=\num{3.85}$.
The timeslice index $\tau$ is integer-valued. The effective mass \meff is given in units of $\kappa$.
\label{fig:plateau_fit}}
\end{figure}

By symmetry, the two Dirac points are indistinguishable, and the correlator expectation values are symmetric in time around 
$\beta/2$ (or $\nt/2$ in units of $\tau=t/\delta$). We therefore average and fold the correlator according to
\begin{equation}
\label{eq:symmetrized}
C(t) := \frac{1}{4} \big( 
C_{KK}^{}(t) + C_{K^\prime K^\prime}(t) + 
C_{KK}^{}(\beta-t) + C_{K^\prime K^\prime}(\beta-t)\big),
\end{equation}
on a configuration-by-configuration basis, 
which increases our numerical precision without the need for generating additional MC configurations. %
In most cases, thermal effects due to backwards-propagating states cannot be completely neglected, and the isolation of an unambiguous exponential decay is difficult.
Instead of obtaining $\Delta$ from
\begin{align}
\label{eq:const fit}
\meff(\tau)\,\delta = \ln\frac{C(\tau)}{C(\tau+1)},
\end{align}
we use the symmetry of the correlator at the Dirac point about $\tau=\beta/2$ and calculate
\begin{equation}
\label{eq:cosh fit}
\meff(\tau)\, \delta = \cosh^{-1}\left(\frac{C(\tau+1)+C(\tau-1)}{2C(\tau)}\right)\ ,
\end{equation}
where $C(\tau)$ is the folded and symmetrized correlator \eqref{eq:symmetrized}.
This hyperbolic cosine form \eqref{eq:cosh fit} is found to be more accurate than the exponential form \eqref{eq:const fit}, 
as the quasiparticle masses are rather small, and the backward-propagating contributions non-negligible (due to the finite extent in $\beta$).
Once the optimal region from which to extract the effective mass has been found (for details, see \Appref{plateau_estimation}), 
we fit the correlator in this region to the form
\begin{align}
C(\tau) = a \cosh\left(\meff\,\delta\left(\tau-N_t/2 \right)\right),
\end{align}
with $a$ and $\meff$ as fit parameters. In \Figref{plateau_fit}, we show the difference between the hyperbolic cosine fit and a direct fit of $\meff(\tau)$. The former is shown
by a blue line and band (statistical error), and the latter by an orange dashed line. While the results are in general agreement, the direct fit of $\meff(\tau)$ tends to overestimate the
gap for small {\effm}es and high statistical noise levels. The estimated systematic error due to the choice of the fit range (as explained in \Appref{plateau_estimation}) is 
shown by the red dot-dashed line. The statistical errors are obtained by a bootstrap procedure, and the total error has been estimated by adding the statistical and
systematic uncertainties in quadrature. The data analysis described in this and the following sections has mostly been performed in the \texttt{R} language~\cite{R_language}, 
in particular using the \texttt{hadron} package~\cite{hadron}.

\subsection{Lattice artifacts \label{sec:lattice_artefacts}}

Once we have determined the single-particle gap $\Delta$, we still need to consider the limits $\delta \to 0$, $L \to \infty$, and $\beta \to \infty$. We shall discuss the
first two limits here, and return to the issue of finite-size scaling in $\beta$ later.

In general, grand canonical MC simulations of the Hubbard model are expected to show very mild finite-size effects which vanish exponentially with $L$, as found by
\Ref{Wang:2017ohn}. This situation is more favourable than in canonical simulations, where observables typically scale as a power-law in $L^{-1}$~\cite{Wang:2017ohn}.
Let us briefly consider the findings of other recent MC studies.
For the extrapolation in $L$, \Refs{Assaad:2013xua,Meng2010} 
(for $\Delta$ and the squared staggered magnetic moment $m_s^2$), \Ref{Otsuka:2015iba} (for $m_s^2$) and \Ref{Buividovich:2018yar} 
(for the square of the total spin per sublattice), find a power-law dependence of the form $a+bL\inverse+cL^{-2}$.
On the other hand, \Ref{Ulybyshev_2017} found little or no dependence on $L$ for the
conductivity.
A side effect of a polynomial dependence on $L^{-1}$ is that (manifestly positive) extrapolated quantities may become
negative in the limit $L \to \infty$.

The continuous time limit $\delta \to 0$ was taken very carefully in \Ref{Otsuka:2015iba},
by simulations at successively smaller $\delta$ until the numerical results stabilized. With the exponential (or compact)
kinetic energy term used in \Ref{Otsuka:2015iba}, the Trotter error of observables should scale as $\ordnung{\delta^2}$. As shown, for instance in 
\Ref{Wynen:2018ryx}, discretization errors of observables for our linearised kinetic energy operator are in general of $\ordnung\delta$. Even with an
exponential kinetic energy operator, some extrapolation in $\delta$ is usually required~\cite{Wynen:2018ryx}. 
For further details concerning the limit $\delta \to 0$, see \Appref{mixed_bias}.

In \Appref{thermal_gap}, we argue that the residual modification of $\Delta$ due to the finite lattice extent $L$ should be proportional to $L^{-3}$.
This is not expected for all observables, but only for those that satisfy two conditions. First, the observable should not (for single MC configurations)
have errors proportional to $L^{-2}$, which are not cancelled by an ensemble average. This condition is satisfied by the correlation functions at the Dirac points, 
but not in general for (squared) magnetic or other locally defined quantities. For example, $m_s^2$ exhibits $\sim L^{-2}$ fluctuations which are 
positive for every MC configuration. As the average of these positive contributions does not vanish, the error is effectively proportional to $L^{-2}$.
Second, the correlation length $\xi$ has to fulfil $\xi \ll L$, such that the error contribution $\sim\exp(-L/\xi)$ as in \Ref{Wang:2017ohn} remains suppressed.
While we expect $L^{-3}$ scaling to hold far from phase transitions, this may break down 
close to criticality, where $\xi \to \infty$. We find that deviations from inverse cubic scaling are small when
\begin{align}
L \gg \frac{\kappa\beta}{\pi},
\end{align}
where $\pi/\beta$ is the minimum Matsubara frequency dominating the correlation length ($\kappa\beta/\pi\gtrsim\xi$). We employed $\kappa \beta\le 12$ implying the requirement $L\gg 4$. Numerically, we find that for $L \ge 9$ our observed dependence on $L$ is entirely governed by inverse cubic scaling (see Figures~\ref{fig:2d_gap_fit} and~\ref{fig:finite_size_always_cubic}).

\subsection{Extrapolation method}

In this study, we have chosen to perform a simultaneous extrapolation in $\delta$ and $L$, by a two-dimensional chi-square minimization.
We use the extrapolation formula
\begin{align}
\mathfrak{O}(L,N_t^{}) =
\mathfrak{O} + a_1^{} L^{-3} + a_2^{} N_t^{-2},
\label{eqn_white_2d}
\end{align}
where $\mathfrak{O}$ is an observable with expected Trotter error of $\ordnung{\delta^2}$, and $a_1, a_2$ are fit parameters.
This is similar to the procedure of \Ref{White:1989zz} for expectation values of the Hamiltonian.
Before we describe our fitting procedure in detail, let us note some advantages of Eq.~(\ref{eqn_white_2d}) relative to a method where observables 
are extracted at fixed $(U,\beta)$ by first taking the temporal continuum limit $\delta \to 0$, 
\begin{align}
\mathfrak{O}(L,N_t^{}) =
\mathfrak{O}(L) + a_2^{}(L) N_t^{-2},
\label{eqn_white_1d_step1}
\end{align}
where each value of $\mathfrak{O}(L)$ and $a_2^{}(L)$ is obtained from a separate chi-square fit, where $L$ is
held fixed. This is followed by
\begin{align}
\mathfrak{O}(L) =
\mathfrak{O} + a_1^{} L^{-3},
\label{eqn_white_1d_step2}
\end{align}
as the final step. Clearly, a two-dimensional chi-square fit using Eq.~(\ref{eqn_white_2d}) involves a much larger
number of degrees of freedom relative to the number of adjustable parameters. This feature makes it easier to judge the quality
of the fit and to identify outliers, which is especially significant when the fit is to be used for extrapolation. While we have presented arguments for the 
expected scaling of $\Delta$ as a function of $L$ and $\delta$, in general the true functional dependence on these variables is not \textit{a priori} known. Thus, the only unbiased check on the 
extrapolation is the quality of each individual fit. This criterion is less stringent, if the data for each value of $L$ is individually extrapolated 
to $\delta \to 0$.

The uncertainties of the fitted parameters have been calculated via 
parametric bootstrap, which means that the bootstrap samples have been generated by drawing from independent normal distributions, defined by every single 
value of $\Delta(L,N_t^{})$ and its individual error.

\subsection{Results \label{sec:gap_results}}

\begin{figure}[t]
\centering
\raisebox{0.016\height}{\includegraphics[width=.45\textwidth]{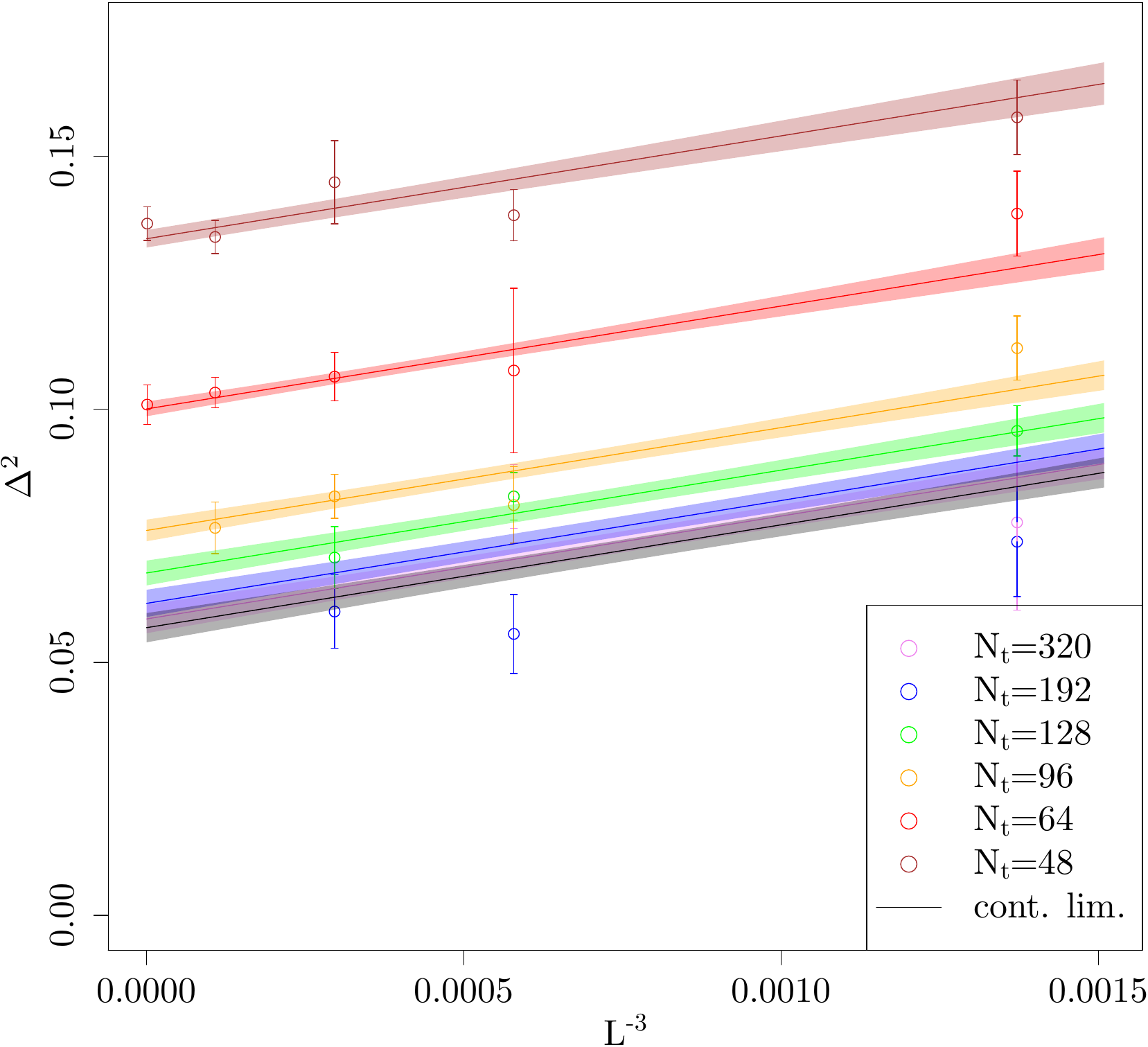}}
\includegraphics[width=.45\textwidth]{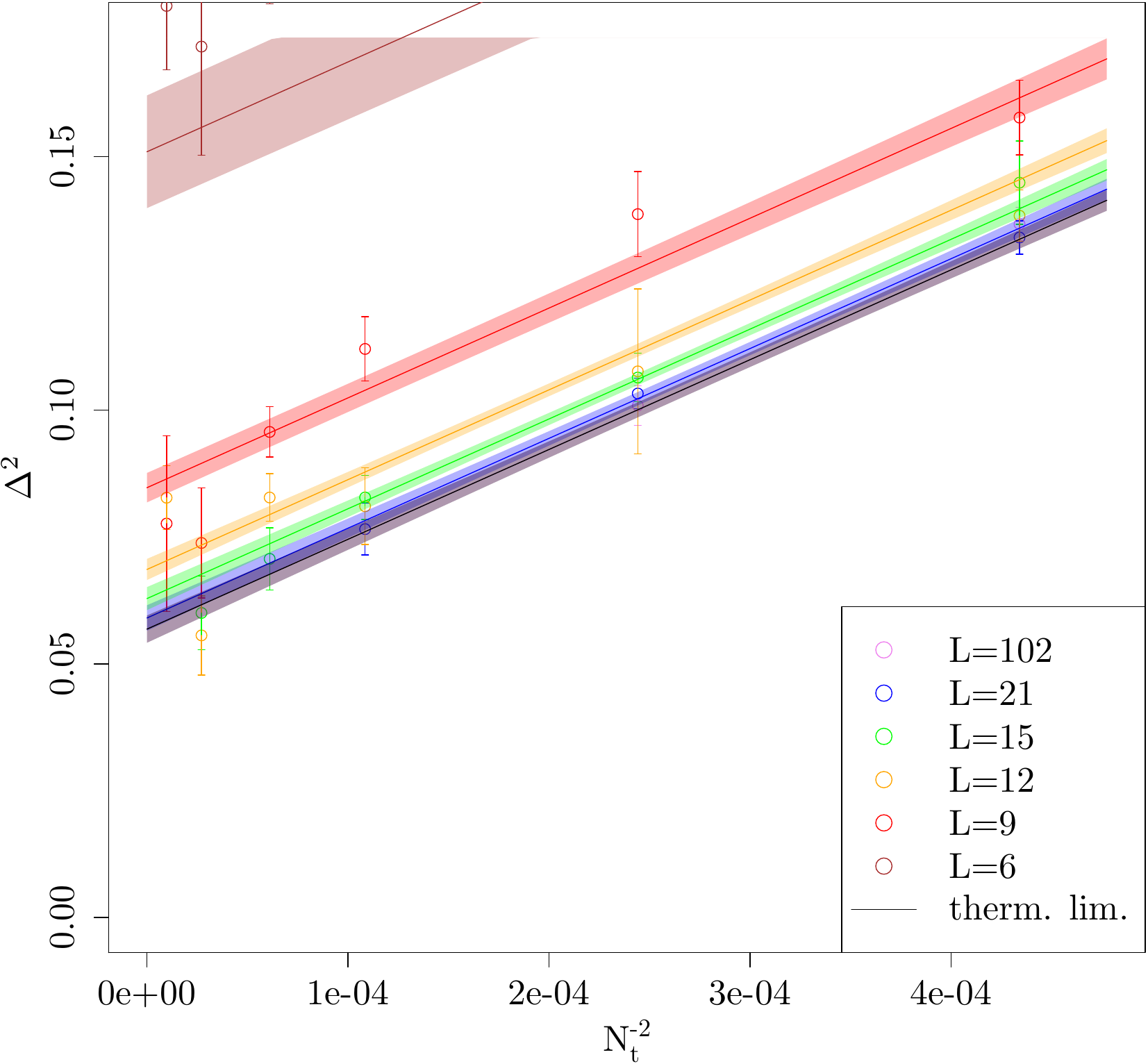}
\caption{Simultaneous two-dimensional fit of $\Delta(N_t,L)$ (in units of $\kappa$) using Eq.~\eqref{eqn_gap_artefacts}, for $\kappa\beta=8$ and $U/\kappa=\num{3.5}$.
Note that Eq.~\eqref{eqn_gap_artefacts} only incorporates effects of $\ordnung{L^{-3}}$ and $\ordnung{N_t^{-2}}$. Data points for $L<9$ have been omitted from the fit, 
but not from the plot. Very small lattices lead to large values of $\Delta$, which are not visible on the scale of the plot.
This fit has $\chi^2/\text{d.o.f.} \simeq \num{1.1}$, corresponding to a p-value of $\simeq \num{0.34}$.
\label{fig:2d_gap_fit}}
\end{figure}

Given the fermion operator described in \Secref{formalism}, we have used Hasenbusch-accelerated HMC~\cite{hasenbusch,2018arXiv180407195K}
to generate a large number of ensembles at different values of $L$, $U/\kappa$, $\kappa\beta$, and $\kappa\delta=\kappa\beta/ N_t$.
We have used six inverse temperatures $\kappa\beta\in\{3, 4, 6, 8, 10, 12\}$\footnote{
These correspond to a highest temperature of $T\approx\SI{1.04e4}{\kelvin}$ and a lowest temperature of $T\approx\SI{2.6e3}{\kelvin}$. According to \Ref{temperature_phase_trans} 
this range is well below the critical temperature $T_c\approx\SI{1.3e4}{\kelvin}$, above which the sublattice symmetry is restored. %
} and a number of couplings in the range $U/\kappa \in [1.0, 5.5]$, which is expected to bracket the critical coupling $U_c/\kappa$ of the AFMI transition. For each temperature and 
coupling, we scanned a large range in $L$ and $\kappa\delta$, to provide reliable extrapolations to the physical limits. These are
$L \in [3, 102]$ \footnote{
	We generated several very large lattices for a few sets of parameters to check the convergence.
	As we observed a nearly flat dependence of $\Delta$ on $L\gtrsim12$, we did most of the analysis with medium-sized lattices ($L\le 21$) in order to conserve
	computation time.
} and $\kappa\delta$ from~$1/4$ down to~$1/40$. Our values of $L$ were chosen to be 0 (mod 3) so that the
non-interacting dispersion has momentum modes exactly at the Dirac points. In other words, we select the lattice geometry to ensure that $E_1-E_0 = 0$ in the absence of
an interaction-induced (Mott) gap. 

Let us briefly summarize our procedure for computing $\Delta$. We have calculated the \spc\ $C(t)$ 
as an expectation value of the inverse fermion matrix, at both independent Dirac-points $K, K^\prime$, and averaged over them to increase our statistics.
For each set of parameters simulated, those correlators have been fitted with an effective mass in order to extract $\Delta$.
We are then in a position to take the limits $\delta \to 0$ and $L \to \infty$, which we accomplish by a simultaneous fit to the functional form
\begin{align}
\Delta^2(L,N_t^{}) =
\Delta_0^2 + c_0^{} N_t^{-2}
+ c_1^{} L^{-3}\,,
\label{eqn_gap_artefacts}
\end{align}
where the fitted quantities are the (infinite-volume, temporal continuum) gap $\Delta_0$, and the leading corrections proportional to $c_0^{}$ and $c_1^{}$.
Around $\delta = 0$, we find the leading $L$-dependence
\begin{align}
\Delta(L) = \Delta_0^{} + \frac{c_1^{}}{2\Delta_0^{}}L^{-3} + \ordnung{L^{-6}},
\end{align}
where $c_1^{}$ is numerically found to be very small. Also, around $L = \infty$, we have
\begin{align}
\Delta(N_t^{}) = \Delta_0^{} + \frac{c_0^{}}{2\Delta_0^{}} N_t^{-2} + \ordnung{N_t^{-4}}\,.
\end{align}
It should be noted that inclusion of a term of order $N_t^{-1}$ did not improve the quality of the fit. This observation 
can be justified by the expected suppression of the linear term due to our mixed differencing scheme. 
Though the term of $\ordnung\delta$ is not removed analytically, it appears small enough to be numerically unresolvable.

\begin{figure}[t]
\begingroup
  \inputencoding{latin1}%
  \makeatletter
  \providecommand\color[2][]{%
    \GenericError{(gnuplot) \space\space\space\@spaces}{%
      Package color not loaded in conjunction with
      terminal option `colourtext'%
    }{See the gnuplot documentation for explanation.%
    }{Either use 'blacktext' in gnuplot or load the package
      color.sty in LaTeX.}%
    \renewcommand\color[2][]{}%
  }%
  \providecommand\includegraphics[2][]{%
    \GenericError{(gnuplot) \space\space\space\@spaces}{%
      Package graphicx or graphics not loaded%
    }{See the gnuplot documentation for explanation.%
    }{The gnuplot epslatex terminal needs graphicx.sty or graphics.sty.}%
    \renewcommand\includegraphics[2][]{}%
  }%
  \providecommand\rotatebox[2]{#2}%
  \@ifundefined{ifGPcolor}{%
    \newif\ifGPcolor
    \GPcolortrue
  }{}%
  \@ifundefined{ifGPblacktext}{%
    \newif\ifGPblacktext
    \GPblacktexttrue
  }{}%
  \let\gplgaddtomacro\g@addto@macro
  \gdef\gplbacktext{}%
  \gdef\gplfronttext{}%
  \makeatother
  \ifGPblacktext
    \def\colorrgb#1{}%
    \def\colorgray#1{}%
  \else
    \ifGPcolor
      \def\colorrgb#1{\color[rgb]{#1}}%
      \def\colorgray#1{\color[gray]{#1}}%
      \expandafter\def\csname LTw\endcsname{\color{white}}%
      \expandafter\def\csname LTb\endcsname{\color{black}}%
      \expandafter\def\csname LTa\endcsname{\color{black}}%
      \expandafter\def\csname LT0\endcsname{\color[rgb]{1,0,0}}%
      \expandafter\def\csname LT1\endcsname{\color[rgb]{0,1,0}}%
      \expandafter\def\csname LT2\endcsname{\color[rgb]{0,0,1}}%
      \expandafter\def\csname LT3\endcsname{\color[rgb]{1,0,1}}%
      \expandafter\def\csname LT4\endcsname{\color[rgb]{0,1,1}}%
      \expandafter\def\csname LT5\endcsname{\color[rgb]{1,1,0}}%
      \expandafter\def\csname LT6\endcsname{\color[rgb]{0,0,0}}%
      \expandafter\def\csname LT7\endcsname{\color[rgb]{1,0.3,0}}%
      \expandafter\def\csname LT8\endcsname{\color[rgb]{0.5,0.5,0.5}}%
    \else
      \def\colorrgb#1{\color{black}}%
      \def\colorgray#1{\color[gray]{#1}}%
      \expandafter\def\csname LTw\endcsname{\color{white}}%
      \expandafter\def\csname LTb\endcsname{\color{black}}%
      \expandafter\def\csname LTa\endcsname{\color{black}}%
      \expandafter\def\csname LT0\endcsname{\color{black}}%
      \expandafter\def\csname LT1\endcsname{\color{black}}%
      \expandafter\def\csname LT2\endcsname{\color{black}}%
      \expandafter\def\csname LT3\endcsname{\color{black}}%
      \expandafter\def\csname LT4\endcsname{\color{black}}%
      \expandafter\def\csname LT5\endcsname{\color{black}}%
      \expandafter\def\csname LT6\endcsname{\color{black}}%
      \expandafter\def\csname LT7\endcsname{\color{black}}%
      \expandafter\def\csname LT8\endcsname{\color{black}}%
    \fi
  \fi
    \setlength{\unitlength}{0.0500bp}%
    \ifx\gptboxheight\undefined%
      \newlength{\gptboxheight}%
      \newlength{\gptboxwidth}%
      \newsavebox{\gptboxtext}%
    \fi%
    \setlength{\fboxrule}{0.5pt}%
    \setlength{\fboxsep}{1pt}%
\begin{picture}(7200.00,5040.00)%
    \gplgaddtomacro\gplbacktext{%
      \csname LTb\endcsname%
      \put(814,921){\makebox(0,0)[r]{\strut{}$0$}}%
      \csname LTb\endcsname%
      \put(814,1354){\makebox(0,0)[r]{\strut{}$0.1$}}%
      \csname LTb\endcsname%
      \put(814,1787){\makebox(0,0)[r]{\strut{}$0.2$}}%
      \csname LTb\endcsname%
      \put(814,2220){\makebox(0,0)[r]{\strut{}$0.3$}}%
      \csname LTb\endcsname%
      \put(814,2653){\makebox(0,0)[r]{\strut{}$0.4$}}%
      \csname LTb\endcsname%
      \put(814,3086){\makebox(0,0)[r]{\strut{}$0.5$}}%
      \csname LTb\endcsname%
      \put(814,3520){\makebox(0,0)[r]{\strut{}$0.6$}}%
      \csname LTb\endcsname%
      \put(814,3953){\makebox(0,0)[r]{\strut{}$0.7$}}%
      \csname LTb\endcsname%
      \put(814,4386){\makebox(0,0)[r]{\strut{}$0.8$}}%
      \csname LTb\endcsname%
      \put(814,4819){\makebox(0,0)[r]{\strut{}$0.9$}}%
      \csname LTb\endcsname%
      \put(1246,484){\makebox(0,0){\strut{}$1$}}%
      \csname LTb\endcsname%
      \put(1997,484){\makebox(0,0){\strut{}$1.5$}}%
      \csname LTb\endcsname%
      \put(2748,484){\makebox(0,0){\strut{}$2$}}%
      \csname LTb\endcsname%
      \put(3499,484){\makebox(0,0){\strut{}$2.5$}}%
      \csname LTb\endcsname%
      \put(4250,484){\makebox(0,0){\strut{}$3$}}%
      \csname LTb\endcsname%
      \put(5001,484){\makebox(0,0){\strut{}$3.5$}}%
      \csname LTb\endcsname%
      \put(5752,484){\makebox(0,0){\strut{}$4$}}%
      \csname LTb\endcsname%
      \put(6503,484){\makebox(0,0){\strut{}$4.5$}}%
    }%
    \gplgaddtomacro\gplfronttext{%
      \csname LTb\endcsname%
      \put(198,2761){\rotatebox{-270}{\makebox(0,0){\strut{}$\Delta_0$}}}%
      \put(3874,154){\makebox(0,0){\strut{}$U$}}%
      \csname LTb\endcsname%
      \put(1870,4646){\makebox(0,0)[r]{\strut{}$\beta = 3$}}%
      \csname LTb\endcsname%
      \put(1870,4426){\makebox(0,0)[r]{\strut{}$\beta = 4$}}%
      \csname LTb\endcsname%
      \put(1870,4206){\makebox(0,0)[r]{\strut{}$\beta = 6$}}%
      \csname LTb\endcsname%
      \put(1870,3986){\makebox(0,0)[r]{\strut{}$\beta = 8$}}%
      \csname LTb\endcsname%
      \put(1870,3766){\makebox(0,0)[r]{\strut{}$\beta = 10$}}%
      \csname LTb\endcsname%
      \put(1870,3546){\makebox(0,0)[r]{\strut{}$\beta = 12$}}%
      \csname LTb\endcsname%
      \put(1870,3326){\makebox(0,0)[r]{\strut{}$\beta = \infty$}}%
    }%
    \gplbacktext
    \put(0,0){\includegraphics{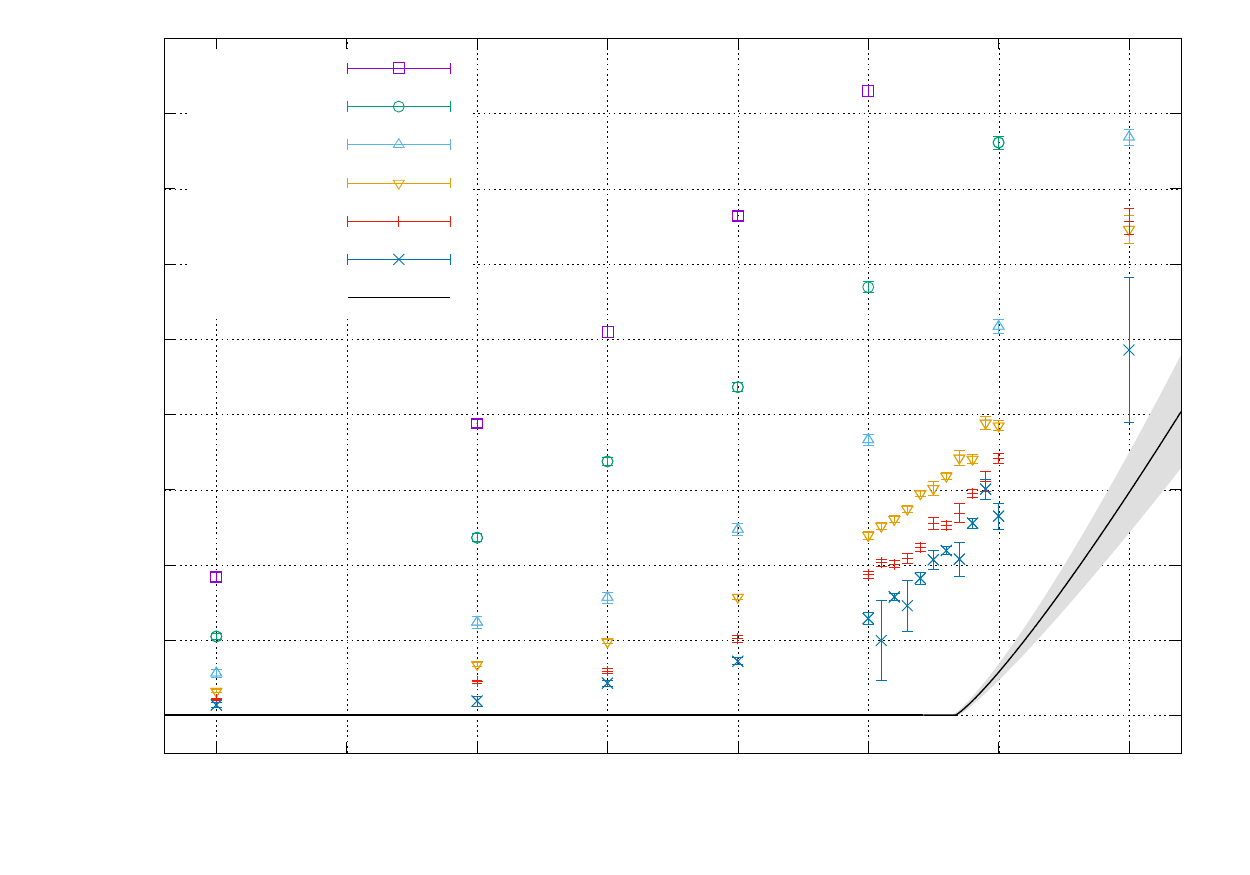}}%
    \gplfronttext
  \end{picture}%
\endgroup
 \caption{The single-particle gap $\Delta_0(U,\beta)$, with all quantities in units of $\kappa$, after the thermodynamic and continuum limit extrapolations.  
We also show $\Delta_0(U,\beta \to \infty)$ as a solid black line with error band (see \Secref{zero temperature extrapolation}). For 
$U< U_c \simeq \critU$ the zero-temperature gap vanishes. 
\label{fig:delta_against_U}}
\end{figure}

An example of such a fit is shown in \Figref{2d_gap_fit}. We find that Eq.~\eqref{eqn_gap_artefacts} describes our data well with only a minimal set of parameters for large $L$ and 
small $\delta$. Lattices with $L\le6$ have been omitted from the fit, as such data does not always lie in the scaling region where the data shows 
$\ordnung{L^{-3}}$ convergence, as explained in \Secref{lattice_artefacts}.
Our results for $\Delta_0$ are shown
in \Figref{delta_against_U} for all values of $U/\kappa$ and $\kappa\beta$, along with an extrapolation (with error band) to zero temperature ($\beta\to\infty$).
For details as to the zero-temperature gap, see \Secref{analysis}.

\section{Analysis \label{sec:analysis}}

We now analyze the single-particle gap $\Delta_0$ as a function of coupling $U-U_c$ and inverse temperature $\beta$, in order to determine the critical coupling
$U_c/\kappa$ of the quantum (AFMI) phase transition, along with some of the associated critical exponents. Our MC results for $\Delta_0$ as a function of $U$ are shown in \Figref{delta_against_U}.
We make use of the standard finite-size scaling (FSS) method~\cite{newmanb99, dutta_aeppli, Shao_2016}, whereby a given observable $\mathfrak{O}$ is described by
\begin{equation}
\mathfrak{O} = (U-U_c)^q \mathcal{F}_{\mathfrak{O}}^{}(L/\xi, \beta/\xi_t^{}), 
\label{eqn:scaling_1}
\end{equation}
in the vicinity of $U_c$, where $q$ is the relevant critical exponent. The scaling function $\mathcal{F}$ accounts for the effects of finite spatial system size $L$ and inverse temperature $\beta$.
The spatial and temporal correlation lengths $\xi$ and $\xi_t$ are
\begin{equation}
\xi \sim (U-U_c)^{-\nu},
\quad
\xi_t^{} \sim (U-U_c)^{-\nu_t^{}} = (U-U_c)^{-z\nu} \sim \xi^z,
\label{eqn:corr}
\end{equation}
such that in the thermodynamic limit we can express the FSS relation as
\begin{equation}
\mathfrak{O} = \beta^{-q/(z\nu)} F_{\mathfrak{O}}^{}(\beta^{1/(z\nu)}(U-U_c)),
\label{eqn:scaling_5}
\end{equation}
which is analogous to \Refs{Assaad:2013xua, Otsuka:2015iba}, apart from the scaling argument here being
$\beta$ instead of $L$, and the correlation length exponent $z\nu$ picked up a factor of the dynamical exponent $z$.

We assume the standard scaling behavior~\cite{Herbut:2009qb} for the single-particle gap 
\begin{equation}
\Delta_0^{} \sim \xi_t^{-1} \sim (U-U_c)^{z\nu},
\end{equation}
as extracted from the asymptotic behavior of the correlator $C(t)$. Therefore, Eq.~\eqref{eqn:scaling_5} gives
\begin{equation}
\Delta_0^{} = \beta^{-1} 
F(\beta^{1/(z\nu)}(U-U_c)),
\label{eqn:scaling_c2}
\end{equation}
as the FSS relation for $\Delta_0$. Our treatment of Eq.~\eqref{eqn:scaling_c2} 
is similar to the data-collapsing procedure of Refs.~\cite{beach2005data,Campostrini:2014lta}.
We define a universal (not explicitly $\beta$-dependent), smooth function $F$ such that
\begin{align}
u & \coloneqq \beta^\mu (U-U_c^{}), \quad \mu:= 1/(z\nu), \\
f & \coloneqq \beta \Delta_0^{},\\
f & \overset{!}{=} F(u),
\label{eq:scaling_for_collapse}
\end{align}
and by adjusting the parameters $\mu$ and $U_c$, we seek to minimize the dependence of the observable $\Delta_0$ on the system size~\cite{19881,Campostrini:2014lta}, in our
case the inverse temperature $\beta$. In practice, these parameters are determined such that all points of the (appropriately scaled) gap $\Delta_0$ lie on a single line in a $u$-$f$ plot.

In this work, we have not computed the AFMI order parameter (staggered magnetization) $m_s$. However, we can make use of the findings of \Ref{Assaad:2013xua} to
provide a first estimate of the associated critical exponent $\beta$ (which we denote by $\tilde\beta$, to avoid confusion with the inverse temperature $\beta$). Specifically, 
\Ref{Assaad:2013xua} found that describing $\Delta_0/U$ according to the FSS relation
\begin{equation}
\Delta_0^{}/U = \beta^{-\tilde\beta/(z\nu)} 
G(\beta^{1/(z\nu)}(U-U_c)),
\label{eqn:scaling_gg}
\end{equation}
produced a scaling function $G$ which was indistinguishable from the true scaling function for $m_s$, in spite of the \textit{ansatz} $m_s \sim \Delta/U$ being a
mean-field result. In our notation, this corresponds to
\begin{align}
g \coloneqq \beta^\zeta \Delta_0^{}/U & \overset{!}{=} G(u), \quad \zeta:= \tilde\beta/(z\nu),
\label{eq:scaling_for_collapse_magn}
\end{align}
similarly to Eq.~\eqref{eq:scaling_for_collapse}.

It should be noted that Refs.~\cite{Toldin:2014sxa, Otsuka:2015iba} also considered sub-leading corrections to the FSS relation for $m_s$.
Here, data points outside of the scaling region have been omitted instead. When taken together with additional fit parameters, 
they do not increase the significance of the fit. On the contrary, they reduce its stability. As a cutoff, we take $\beta U < 8$ for Eq.~\eqref{eq:scaling_for_collapse} and $\beta U < 10$ for 
Eq.~\eqref{eq:scaling_for_collapse_magn}. 
The reason for this limited scaling region can be understood as follows. The gap vanishes at $U=0$ regardless of $\beta$, thus $\Delta_0(u=-\beta^\mu U_c)=0$, and therefore data points with small 
$\beta$ and $U$ no longer collapse onto $F(u)$. Similarly, as we show in Appendix~\ref{sec:thermal_gap}, for $U \ll U_c$ we have $\Delta_0\sim U\beta^{-2}$, which implies $g\sim\beta^{\zeta-2}\approx\beta^{-1}$, and represents a strong deviation from the expected scaling. With decreasing $\beta$, the effect increases and materialises at larger $U$.
We shall revisit the issue of sub-leading corrections in a future MC study of $m_s$.

\begin{figure}[ht]
\begingroup
  \inputencoding{latin1}%
  \makeatletter
  \providecommand\color[2][]{%
    \GenericError{(gnuplot) \space\space\space\@spaces}{%
      Package color not loaded in conjunction with
      terminal option `colourtext'%
    }{See the gnuplot documentation for explanation.%
    }{Either use 'blacktext' in gnuplot or load the package
      color.sty in LaTeX.}%
    \renewcommand\color[2][]{}%
  }%
  \providecommand\includegraphics[2][]{%
    \GenericError{(gnuplot) \space\space\space\@spaces}{%
      Package graphicx or graphics not loaded%
    }{See the gnuplot documentation for explanation.%
    }{The gnuplot epslatex terminal needs graphicx.sty or graphics.sty.}%
    \renewcommand\includegraphics[2][]{}%
  }%
  \providecommand\rotatebox[2]{#2}%
  \@ifundefined{ifGPcolor}{%
    \newif\ifGPcolor
    \GPcolortrue
  }{}%
  \@ifundefined{ifGPblacktext}{%
    \newif\ifGPblacktext
    \GPblacktexttrue
  }{}%
  \let\gplgaddtomacro\g@addto@macro
  \gdef\gplbacktext{}%
  \gdef\gplfronttext{}%
  \makeatother
  \ifGPblacktext
    \def\colorrgb#1{}%
    \def\colorgray#1{}%
  \else
    \ifGPcolor
      \def\colorrgb#1{\color[rgb]{#1}}%
      \def\colorgray#1{\color[gray]{#1}}%
      \expandafter\def\csname LTw\endcsname{\color{white}}%
      \expandafter\def\csname LTb\endcsname{\color{black}}%
      \expandafter\def\csname LTa\endcsname{\color{black}}%
      \expandafter\def\csname LT0\endcsname{\color[rgb]{1,0,0}}%
      \expandafter\def\csname LT1\endcsname{\color[rgb]{0,1,0}}%
      \expandafter\def\csname LT2\endcsname{\color[rgb]{0,0,1}}%
      \expandafter\def\csname LT3\endcsname{\color[rgb]{1,0,1}}%
      \expandafter\def\csname LT4\endcsname{\color[rgb]{0,1,1}}%
      \expandafter\def\csname LT5\endcsname{\color[rgb]{1,1,0}}%
      \expandafter\def\csname LT6\endcsname{\color[rgb]{0,0,0}}%
      \expandafter\def\csname LT7\endcsname{\color[rgb]{1,0.3,0}}%
      \expandafter\def\csname LT8\endcsname{\color[rgb]{0.5,0.5,0.5}}%
    \else
      \def\colorrgb#1{\color{black}}%
      \def\colorgray#1{\color[gray]{#1}}%
      \expandafter\def\csname LTw\endcsname{\color{white}}%
      \expandafter\def\csname LTb\endcsname{\color{black}}%
      \expandafter\def\csname LTa\endcsname{\color{black}}%
      \expandafter\def\csname LT0\endcsname{\color{black}}%
      \expandafter\def\csname LT1\endcsname{\color{black}}%
      \expandafter\def\csname LT2\endcsname{\color{black}}%
      \expandafter\def\csname LT3\endcsname{\color{black}}%
      \expandafter\def\csname LT4\endcsname{\color{black}}%
      \expandafter\def\csname LT5\endcsname{\color{black}}%
      \expandafter\def\csname LT6\endcsname{\color{black}}%
      \expandafter\def\csname LT7\endcsname{\color{black}}%
      \expandafter\def\csname LT8\endcsname{\color{black}}%
    \fi
  \fi
    \setlength{\unitlength}{0.0500bp}%
    \ifx\gptboxheight\undefined%
      \newlength{\gptboxheight}%
      \newlength{\gptboxwidth}%
      \newsavebox{\gptboxtext}%
    \fi%
    \setlength{\fboxrule}{0.5pt}%
    \setlength{\fboxsep}{1pt}%
\begin{picture}(4896.00,3772.00)%
    \gplgaddtomacro\gplbacktext{%
      \csname LTb\endcsname%
      \put(682,704){\makebox(0,0)[r]{\strut{}$0$}}%
      \csname LTb\endcsname%
      \put(682,1179){\makebox(0,0)[r]{\strut{}$2$}}%
      \csname LTb\endcsname%
      \put(682,1653){\makebox(0,0)[r]{\strut{}$4$}}%
      \csname LTb\endcsname%
      \put(682,2128){\makebox(0,0)[r]{\strut{}$6$}}%
      \csname LTb\endcsname%
      \put(682,2602){\makebox(0,0)[r]{\strut{}$8$}}%
      \csname LTb\endcsname%
      \put(682,3077){\makebox(0,0)[r]{\strut{}$10$}}%
      \csname LTb\endcsname%
      \put(682,3551){\makebox(0,0)[r]{\strut{}$12$}}%
      \csname LTb\endcsname%
      \put(814,484){\makebox(0,0){\strut{}$-25$}}%
      \csname LTb\endcsname%
      \put(1340,484){\makebox(0,0){\strut{}$-20$}}%
      \csname LTb\endcsname%
      \put(1867,484){\makebox(0,0){\strut{}$-15$}}%
      \csname LTb\endcsname%
      \put(2393,484){\makebox(0,0){\strut{}$-10$}}%
      \csname LTb\endcsname%
      \put(2920,484){\makebox(0,0){\strut{}$-5$}}%
      \csname LTb\endcsname%
      \put(3446,484){\makebox(0,0){\strut{}$0$}}%
      \csname LTb\endcsname%
      \put(3973,484){\makebox(0,0){\strut{}$5$}}%
      \csname LTb\endcsname%
      \put(4499,484){\makebox(0,0){\strut{}$10$}}%
    }%
    \gplgaddtomacro\gplfronttext{%
      \csname LTb\endcsname%
      \put(198,2127){\rotatebox{-270}{\makebox(0,0){\strut{}$\beta\Delta_0$}}}%
      \put(2656,154){\makebox(0,0){\strut{}$\beta^\mu(U-U_c)$}}%
      \csname LTb\endcsname%
      \put(1738,3378){\makebox(0,0)[r]{\strut{}$\beta = 3$}}%
      \csname LTb\endcsname%
      \put(1738,3158){\makebox(0,0)[r]{\strut{}$\beta = 4$}}%
      \csname LTb\endcsname%
      \put(1738,2938){\makebox(0,0)[r]{\strut{}$\beta = 6$}}%
      \csname LTb\endcsname%
      \put(1738,2718){\makebox(0,0)[r]{\strut{}$\beta = 8$}}%
      \csname LTb\endcsname%
      \put(1738,2498){\makebox(0,0)[r]{\strut{}$\beta = 10$}}%
      \csname LTb\endcsname%
      \put(1738,2278){\makebox(0,0)[r]{\strut{}$\beta = 12$}}%
    }%
    \gplbacktext
    \put(0,0){\includegraphics{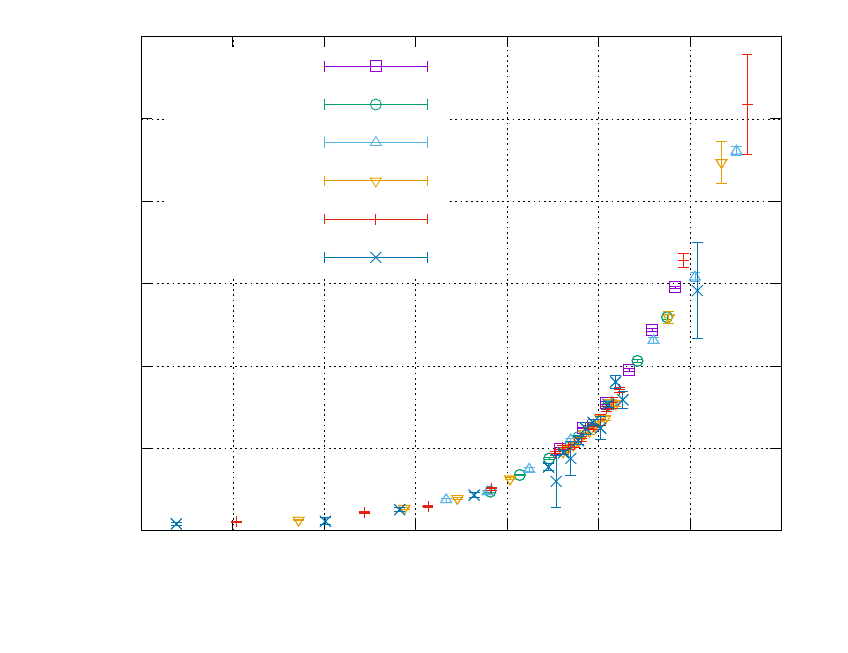}}%
    \gplfronttext
  \end{picture}%
\endgroup
\begingroup
  \inputencoding{latin1}%
  \makeatletter
  \providecommand\color[2][]{%
    \GenericError{(gnuplot) \space\space\space\@spaces}{%
      Package color not loaded in conjunction with
      terminal option `colourtext'%
    }{See the gnuplot documentation for explanation.%
    }{Either use 'blacktext' in gnuplot or load the package
      color.sty in LaTeX.}%
    \renewcommand\color[2][]{}%
  }%
  \providecommand\includegraphics[2][]{%
    \GenericError{(gnuplot) \space\space\space\@spaces}{%
      Package graphicx or graphics not loaded%
    }{See the gnuplot documentation for explanation.%
    }{The gnuplot epslatex terminal needs graphicx.sty or graphics.sty.}%
    \renewcommand\includegraphics[2][]{}%
  }%
  \providecommand\rotatebox[2]{#2}%
  \@ifundefined{ifGPcolor}{%
    \newif\ifGPcolor
    \GPcolortrue
  }{}%
  \@ifundefined{ifGPblacktext}{%
    \newif\ifGPblacktext
    \GPblacktexttrue
  }{}%
  \let\gplgaddtomacro\g@addto@macro
  \gdef\gplbacktext{}%
  \gdef\gplfronttext{}%
  \makeatother
  \ifGPblacktext
    \def\colorrgb#1{}%
    \def\colorgray#1{}%
  \else
    \ifGPcolor
      \def\colorrgb#1{\color[rgb]{#1}}%
      \def\colorgray#1{\color[gray]{#1}}%
      \expandafter\def\csname LTw\endcsname{\color{white}}%
      \expandafter\def\csname LTb\endcsname{\color{black}}%
      \expandafter\def\csname LTa\endcsname{\color{black}}%
      \expandafter\def\csname LT0\endcsname{\color[rgb]{1,0,0}}%
      \expandafter\def\csname LT1\endcsname{\color[rgb]{0,1,0}}%
      \expandafter\def\csname LT2\endcsname{\color[rgb]{0,0,1}}%
      \expandafter\def\csname LT3\endcsname{\color[rgb]{1,0,1}}%
      \expandafter\def\csname LT4\endcsname{\color[rgb]{0,1,1}}%
      \expandafter\def\csname LT5\endcsname{\color[rgb]{1,1,0}}%
      \expandafter\def\csname LT6\endcsname{\color[rgb]{0,0,0}}%
      \expandafter\def\csname LT7\endcsname{\color[rgb]{1,0.3,0}}%
      \expandafter\def\csname LT8\endcsname{\color[rgb]{0.5,0.5,0.5}}%
    \else
      \def\colorrgb#1{\color{black}}%
      \def\colorgray#1{\color[gray]{#1}}%
      \expandafter\def\csname LTw\endcsname{\color{white}}%
      \expandafter\def\csname LTb\endcsname{\color{black}}%
      \expandafter\def\csname LTa\endcsname{\color{black}}%
      \expandafter\def\csname LT0\endcsname{\color{black}}%
      \expandafter\def\csname LT1\endcsname{\color{black}}%
      \expandafter\def\csname LT2\endcsname{\color{black}}%
      \expandafter\def\csname LT3\endcsname{\color{black}}%
      \expandafter\def\csname LT4\endcsname{\color{black}}%
      \expandafter\def\csname LT5\endcsname{\color{black}}%
      \expandafter\def\csname LT6\endcsname{\color{black}}%
      \expandafter\def\csname LT7\endcsname{\color{black}}%
      \expandafter\def\csname LT8\endcsname{\color{black}}%
    \fi
  \fi
    \setlength{\unitlength}{0.0500bp}%
    \ifx\gptboxheight\undefined%
      \newlength{\gptboxheight}%
      \newlength{\gptboxwidth}%
      \newsavebox{\gptboxtext}%
    \fi%
    \setlength{\fboxrule}{0.5pt}%
    \setlength{\fboxsep}{1pt}%
\begin{picture}(4896.00,3772.00)%
    \gplgaddtomacro\gplbacktext{%
      \csname LTb\endcsname%
      \put(814,704){\makebox(0,0)[r]{\strut{}$0$}}%
      \csname LTb\endcsname%
      \put(814,989){\makebox(0,0)[r]{\strut{}$0.2$}}%
      \csname LTb\endcsname%
      \put(814,1273){\makebox(0,0)[r]{\strut{}$0.4$}}%
      \csname LTb\endcsname%
      \put(814,1558){\makebox(0,0)[r]{\strut{}$0.6$}}%
      \csname LTb\endcsname%
      \put(814,1843){\makebox(0,0)[r]{\strut{}$0.8$}}%
      \csname LTb\endcsname%
      \put(814,2128){\makebox(0,0)[r]{\strut{}$1$}}%
      \csname LTb\endcsname%
      \put(814,2412){\makebox(0,0)[r]{\strut{}$1.2$}}%
      \csname LTb\endcsname%
      \put(814,2697){\makebox(0,0)[r]{\strut{}$1.4$}}%
      \csname LTb\endcsname%
      \put(814,2982){\makebox(0,0)[r]{\strut{}$1.6$}}%
      \csname LTb\endcsname%
      \put(814,3266){\makebox(0,0)[r]{\strut{}$1.8$}}%
      \csname LTb\endcsname%
      \put(814,3551){\makebox(0,0)[r]{\strut{}$2$}}%
      \csname LTb\endcsname%
      \put(946,484){\makebox(0,0){\strut{}$-25$}}%
      \csname LTb\endcsname%
      \put(1454,484){\makebox(0,0){\strut{}$-20$}}%
      \csname LTb\endcsname%
      \put(1961,484){\makebox(0,0){\strut{}$-15$}}%
      \csname LTb\endcsname%
      \put(2469,484){\makebox(0,0){\strut{}$-10$}}%
      \csname LTb\endcsname%
      \put(2976,484){\makebox(0,0){\strut{}$-5$}}%
      \csname LTb\endcsname%
      \put(3484,484){\makebox(0,0){\strut{}$0$}}%
      \csname LTb\endcsname%
      \put(3991,484){\makebox(0,0){\strut{}$5$}}%
      \csname LTb\endcsname%
      \put(4499,484){\makebox(0,0){\strut{}$10$}}%
    }%
    \gplgaddtomacro\gplfronttext{%
      \csname LTb\endcsname%
      \put(198,2127){\rotatebox{-270}{\makebox(0,0){\strut{}$\beta^\zeta\Delta_0/U$}}}%
      \put(2722,154){\makebox(0,0){\strut{}$\beta^\mu(U-U_c)$}}%
      \csname LTb\endcsname%
      \put(1870,3378){\makebox(0,0)[r]{\strut{}$\beta = 3$}}%
      \csname LTb\endcsname%
      \put(1870,3158){\makebox(0,0)[r]{\strut{}$\beta = 4$}}%
      \csname LTb\endcsname%
      \put(1870,2938){\makebox(0,0)[r]{\strut{}$\beta = 6$}}%
      \csname LTb\endcsname%
      \put(1870,2718){\makebox(0,0)[r]{\strut{}$\beta = 8$}}%
      \csname LTb\endcsname%
      \put(1870,2498){\makebox(0,0)[r]{\strut{}$\beta = 10$}}%
      \csname LTb\endcsname%
      \put(1870,2278){\makebox(0,0)[r]{\strut{}$\beta = 12$}}%
    }%
    \gplbacktext
    \put(0,0){\includegraphics{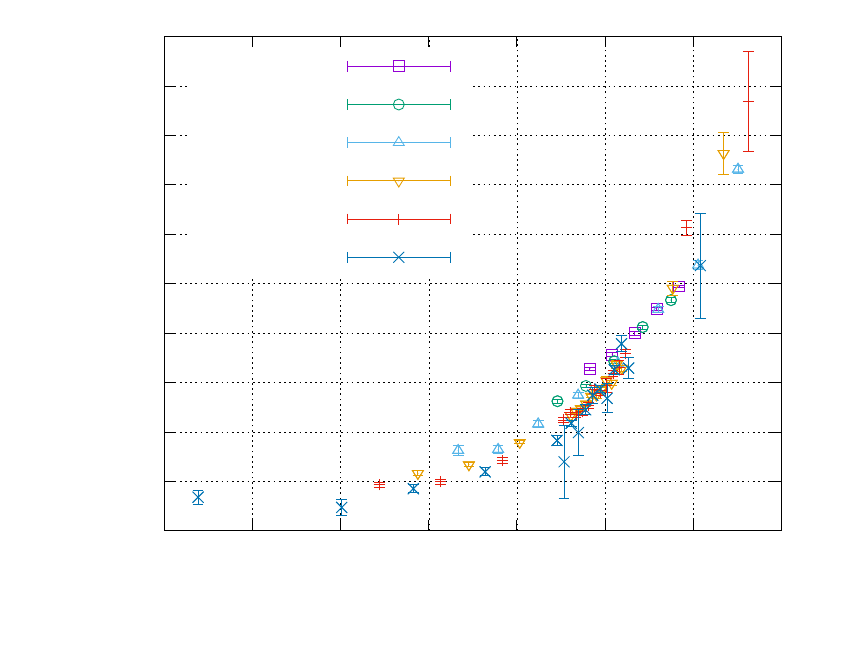}}%
    \gplfronttext
  \end{picture}%
\endgroup
 	\caption{$f = \beta\Delta_0$ (left panel) and $g = \beta^\zeta\Delta_0/U$ (right panel) as a function of $u = \beta^\mu(U-U_c)$ according to Eqs.~\eqref{eq:scaling_for_collapse} and~\eqref{eq:scaling_for_collapse_magn}, respectively. All quantities are plotted with the parameters for optimal data collapse, and expressed in units of $\kappa$.}
	\label{fig:fully_collapsed_data}
\end{figure}

We perform the data-collapse analysis by first interpolating the data using a smoothed cubic spline for each
value of $\beta$. Next we calculate the squared differences between the interpolations for each pair of two different temperatures and integrate over these squared differences. Last the sum over all the integrals is minimised. The resulting optimal data collapse plots are shown 
in \Figref{fully_collapsed_data}. The optimization of Eq.~\eqref{eq:scaling_for_collapse} is performed first (see \Figref{fully_collapsed_data}, left panel). This 
yields $\mu=\critMu$, therefore $z\nu=\critNu$, and $U_c/\kappa=\critU$, where the errors have been determined using parametric bootstrap. 
Second, the data collapse is performed for Eq.~\eqref{eq:scaling_for_collapse_magn} (see \Figref{fully_collapsed_data}, right panel) 
which gives $\zeta=\critZeta$, such that $\tilde\beta=\critExp$. While the $u$-$f$ collapse is satisfactory, the $u$-$g$ collapse does not 
materialize for $u \ll 0$. This is not surprising, because our scaling \textit{ansatz} does not account for the thermal gap, which we discuss in \Appref{thermal_gap}. 
Moreover, these findings are consistent with the notion that the AFMI state predicted by mean-field theory is destroyed by quantum fluctuations at weak coupling, at which point
the relation $m_s \sim \Delta/U$ ceases to be valid.
Also, note that in \Ref{Assaad:2013xua} the plot corresponding to our \Figref{fully_collapsed_data} starts at $u=-2$, a region where our $u$-$g$ collapse works 
out as well (for sufficiently large $\beta$).

We briefly describe our method for determining the errors of the fitted quantities and their correlation matrix. We use a mixed error propagation scheme, consisting of a parametric bootstrap part 
as before, and an additional influence due to a direct propagation of the bootstrap samples of $\mu$ and $U_c$. For every $(\mu,U_c)$-sample we generate a new data set, following the 
parametric sampling system. Then, the variation of the composite samples should mirror the total uncertainty of the results.
The correlation matrix\footnote{We give enough digits to yield percent-level matching to our full numerical results when inverting the correlation matrix.}
\begin{align}
	\mathrm{corr}(U_c,\, z\nu,\, \tilde\beta) &= \matr{S[round-mode=places,round-precision = 4]S[round-mode=places,round-precision = 4]S[round-mode=places,round-precision = 4]}{
	 1.00000000 & -0.01987849 & -0.2330682\\
	-0.01987849 &  1.00000000 &  0.8649537\\
	-0.23306820 &  0.86495372 &  1.0000000
		}
\end{align}
clearly shows a strong correlation between $z\nu$ and $\tilde\beta$ (due to the influence of $\mu$ on both of them), whereas $U_c$ is weakly correlated.

\begin{figure}[ht]
\begingroup
  \inputencoding{latin1}%
  \makeatletter
  \providecommand\color[2][]{%
    \GenericError{(gnuplot) \space\space\space\@spaces}{%
      Package color not loaded in conjunction with
      terminal option `colourtext'%
    }{See the gnuplot documentation for explanation.%
    }{Either use 'blacktext' in gnuplot or load the package
      color.sty in LaTeX.}%
    \renewcommand\color[2][]{}%
  }%
  \providecommand\includegraphics[2][]{%
    \GenericError{(gnuplot) \space\space\space\@spaces}{%
      Package graphicx or graphics not loaded%
    }{See the gnuplot documentation for explanation.%
    }{The gnuplot epslatex terminal needs graphicx.sty or graphics.sty.}%
    \renewcommand\includegraphics[2][]{}%
  }%
  \providecommand\rotatebox[2]{#2}%
  \@ifundefined{ifGPcolor}{%
    \newif\ifGPcolor
    \GPcolortrue
  }{}%
  \@ifundefined{ifGPblacktext}{%
    \newif\ifGPblacktext
    \GPblacktexttrue
  }{}%
  \let\gplgaddtomacro\g@addto@macro
  \gdef\gplbacktext{}%
  \gdef\gplfronttext{}%
  \makeatother
  \ifGPblacktext
    \def\colorrgb#1{}%
    \def\colorgray#1{}%
  \else
    \ifGPcolor
      \def\colorrgb#1{\color[rgb]{#1}}%
      \def\colorgray#1{\color[gray]{#1}}%
      \expandafter\def\csname LTw\endcsname{\color{white}}%
      \expandafter\def\csname LTb\endcsname{\color{black}}%
      \expandafter\def\csname LTa\endcsname{\color{black}}%
      \expandafter\def\csname LT0\endcsname{\color[rgb]{1,0,0}}%
      \expandafter\def\csname LT1\endcsname{\color[rgb]{0,1,0}}%
      \expandafter\def\csname LT2\endcsname{\color[rgb]{0,0,1}}%
      \expandafter\def\csname LT3\endcsname{\color[rgb]{1,0,1}}%
      \expandafter\def\csname LT4\endcsname{\color[rgb]{0,1,1}}%
      \expandafter\def\csname LT5\endcsname{\color[rgb]{1,1,0}}%
      \expandafter\def\csname LT6\endcsname{\color[rgb]{0,0,0}}%
      \expandafter\def\csname LT7\endcsname{\color[rgb]{1,0.3,0}}%
      \expandafter\def\csname LT8\endcsname{\color[rgb]{0.5,0.5,0.5}}%
    \else
      \def\colorrgb#1{\color{black}}%
      \def\colorgray#1{\color[gray]{#1}}%
      \expandafter\def\csname LTw\endcsname{\color{white}}%
      \expandafter\def\csname LTb\endcsname{\color{black}}%
      \expandafter\def\csname LTa\endcsname{\color{black}}%
      \expandafter\def\csname LT0\endcsname{\color{black}}%
      \expandafter\def\csname LT1\endcsname{\color{black}}%
      \expandafter\def\csname LT2\endcsname{\color{black}}%
      \expandafter\def\csname LT3\endcsname{\color{black}}%
      \expandafter\def\csname LT4\endcsname{\color{black}}%
      \expandafter\def\csname LT5\endcsname{\color{black}}%
      \expandafter\def\csname LT6\endcsname{\color{black}}%
      \expandafter\def\csname LT7\endcsname{\color{black}}%
      \expandafter\def\csname LT8\endcsname{\color{black}}%
    \fi
  \fi
    \setlength{\unitlength}{0.0500bp}%
    \ifx\gptboxheight\undefined%
      \newlength{\gptboxheight}%
      \newlength{\gptboxwidth}%
      \newsavebox{\gptboxtext}%
    \fi%
    \setlength{\fboxrule}{0.5pt}%
    \setlength{\fboxsep}{1pt}%
\begin{picture}(4896.00,3772.00)%
    \gplgaddtomacro\gplbacktext{%
      \csname LTb\endcsname%
      \put(550,704){\makebox(0,0)[r]{\strut{}$0$}}%
      \csname LTb\endcsname%
      \put(550,1111){\makebox(0,0)[r]{\strut{}$1$}}%
      \csname LTb\endcsname%
      \put(550,1517){\makebox(0,0)[r]{\strut{}$2$}}%
      \csname LTb\endcsname%
      \put(550,1924){\makebox(0,0)[r]{\strut{}$3$}}%
      \csname LTb\endcsname%
      \put(550,2331){\makebox(0,0)[r]{\strut{}$4$}}%
      \csname LTb\endcsname%
      \put(550,2738){\makebox(0,0)[r]{\strut{}$5$}}%
      \csname LTb\endcsname%
      \put(550,3144){\makebox(0,0)[r]{\strut{}$6$}}%
      \csname LTb\endcsname%
      \put(550,3551){\makebox(0,0)[r]{\strut{}$7$}}%
      \csname LTb\endcsname%
      \put(682,484){\makebox(0,0){\strut{}$2.8$}}%
      \csname LTb\endcsname%
      \put(1084,484){\makebox(0,0){\strut{}$3$}}%
      \csname LTb\endcsname%
      \put(1486,484){\makebox(0,0){\strut{}$3.2$}}%
      \csname LTb\endcsname%
      \put(1887,484){\makebox(0,0){\strut{}$3.4$}}%
      \csname LTb\endcsname%
      \put(2289,484){\makebox(0,0){\strut{}$3.6$}}%
      \csname LTb\endcsname%
      \put(2846,484){\makebox(0,0){\strut{}$U_c^f$}}%
      \csname LTb\endcsname%
      \put(3093,484){\makebox(0,0){\strut{}$4$}}%
      \csname LTb\endcsname%
      \put(3495,484){\makebox(0,0){\strut{}$4.2$}}%
      \csname LTb\endcsname%
      \put(3896,484){\makebox(0,0){\strut{}$4.4$}}%
      \csname LTb\endcsname%
      \put(4298,484){\makebox(0,0){\strut{}$4.6$}}%
    }%
    \gplgaddtomacro\gplfronttext{%
      \csname LTb\endcsname%
      \put(198,2127){\rotatebox{-270}{\makebox(0,0){\strut{}$\beta\Delta_0$}}}%
      \put(2590,154){\makebox(0,0){\strut{}$U$}}%
      \csname LTb\endcsname%
      \put(1606,3378){\makebox(0,0)[r]{\strut{}$\beta = 3$}}%
      \csname LTb\endcsname%
      \put(1606,3158){\makebox(0,0)[r]{\strut{}$\beta = 4$}}%
      \csname LTb\endcsname%
      \put(1606,2938){\makebox(0,0)[r]{\strut{}$\beta = 6$}}%
      \csname LTb\endcsname%
      \put(1606,2718){\makebox(0,0)[r]{\strut{}$\beta = 8$}}%
      \csname LTb\endcsname%
      \put(1606,2498){\makebox(0,0)[r]{\strut{}$\beta = 10$}}%
      \csname LTb\endcsname%
      \put(1606,2278){\makebox(0,0)[r]{\strut{}$\beta = 12$}}%
    }%
    \gplbacktext
    \put(0,0){\includegraphics{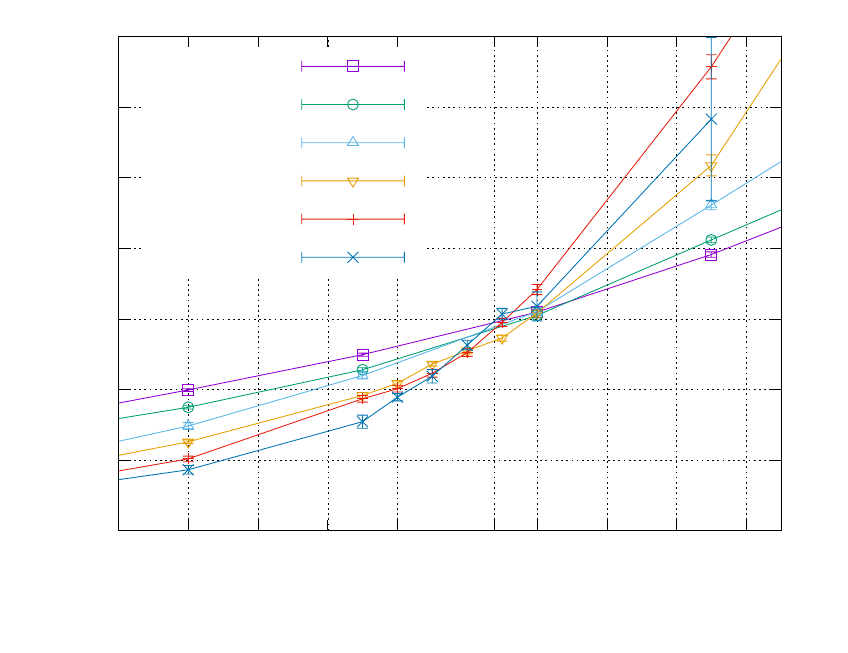}}%
    \gplfronttext
  \end{picture}%
\endgroup
\begingroup
  \inputencoding{latin1}%
  \makeatletter
  \providecommand\color[2][]{%
    \GenericError{(gnuplot) \space\space\space\@spaces}{%
      Package color not loaded in conjunction with
      terminal option `colourtext'%
    }{See the gnuplot documentation for explanation.%
    }{Either use 'blacktext' in gnuplot or load the package
      color.sty in LaTeX.}%
    \renewcommand\color[2][]{}%
  }%
  \providecommand\includegraphics[2][]{%
    \GenericError{(gnuplot) \space\space\space\@spaces}{%
      Package graphicx or graphics not loaded%
    }{See the gnuplot documentation for explanation.%
    }{The gnuplot epslatex terminal needs graphicx.sty or graphics.sty.}%
    \renewcommand\includegraphics[2][]{}%
  }%
  \providecommand\rotatebox[2]{#2}%
  \@ifundefined{ifGPcolor}{%
    \newif\ifGPcolor
    \GPcolortrue
  }{}%
  \@ifundefined{ifGPblacktext}{%
    \newif\ifGPblacktext
    \GPblacktexttrue
  }{}%
  \let\gplgaddtomacro\g@addto@macro
  \gdef\gplbacktext{}%
  \gdef\gplfronttext{}%
  \makeatother
  \ifGPblacktext
    \def\colorrgb#1{}%
    \def\colorgray#1{}%
  \else
    \ifGPcolor
      \def\colorrgb#1{\color[rgb]{#1}}%
      \def\colorgray#1{\color[gray]{#1}}%
      \expandafter\def\csname LTw\endcsname{\color{white}}%
      \expandafter\def\csname LTb\endcsname{\color{black}}%
      \expandafter\def\csname LTa\endcsname{\color{black}}%
      \expandafter\def\csname LT0\endcsname{\color[rgb]{1,0,0}}%
      \expandafter\def\csname LT1\endcsname{\color[rgb]{0,1,0}}%
      \expandafter\def\csname LT2\endcsname{\color[rgb]{0,0,1}}%
      \expandafter\def\csname LT3\endcsname{\color[rgb]{1,0,1}}%
      \expandafter\def\csname LT4\endcsname{\color[rgb]{0,1,1}}%
      \expandafter\def\csname LT5\endcsname{\color[rgb]{1,1,0}}%
      \expandafter\def\csname LT6\endcsname{\color[rgb]{0,0,0}}%
      \expandafter\def\csname LT7\endcsname{\color[rgb]{1,0.3,0}}%
      \expandafter\def\csname LT8\endcsname{\color[rgb]{0.5,0.5,0.5}}%
    \else
      \def\colorrgb#1{\color{black}}%
      \def\colorgray#1{\color[gray]{#1}}%
      \expandafter\def\csname LTw\endcsname{\color{white}}%
      \expandafter\def\csname LTb\endcsname{\color{black}}%
      \expandafter\def\csname LTa\endcsname{\color{black}}%
      \expandafter\def\csname LT0\endcsname{\color{black}}%
      \expandafter\def\csname LT1\endcsname{\color{black}}%
      \expandafter\def\csname LT2\endcsname{\color{black}}%
      \expandafter\def\csname LT3\endcsname{\color{black}}%
      \expandafter\def\csname LT4\endcsname{\color{black}}%
      \expandafter\def\csname LT5\endcsname{\color{black}}%
      \expandafter\def\csname LT6\endcsname{\color{black}}%
      \expandafter\def\csname LT7\endcsname{\color{black}}%
      \expandafter\def\csname LT8\endcsname{\color{black}}%
    \fi
  \fi
    \setlength{\unitlength}{0.0500bp}%
    \ifx\gptboxheight\undefined%
      \newlength{\gptboxheight}%
      \newlength{\gptboxwidth}%
      \newsavebox{\gptboxtext}%
    \fi%
    \setlength{\fboxrule}{0.5pt}%
    \setlength{\fboxsep}{1pt}%
\begin{picture}(4896.00,3772.00)%
    \gplgaddtomacro\gplbacktext{%
      \csname LTb\endcsname%
      \put(814,704){\makebox(0,0)[r]{\strut{}$0.2$}}%
      \csname LTb\endcsname%
      \put(814,1179){\makebox(0,0)[r]{\strut{}$0.4$}}%
      \csname LTb\endcsname%
      \put(814,1653){\makebox(0,0)[r]{\strut{}$0.6$}}%
      \csname LTb\endcsname%
      \put(814,2128){\makebox(0,0)[r]{\strut{}$0.8$}}%
      \csname LTb\endcsname%
      \put(814,2602){\makebox(0,0)[r]{\strut{}$1$}}%
      \csname LTb\endcsname%
      \put(814,3076){\makebox(0,0)[r]{\strut{}$1.2$}}%
      \csname LTb\endcsname%
      \put(814,3551){\makebox(0,0)[r]{\strut{}$1.4$}}%
      \csname LTb\endcsname%
      \put(946,484){\makebox(0,0){\strut{}$2.8$}}%
      \csname LTb\endcsname%
      \put(1320,484){\makebox(0,0){\strut{}$3$}}%
      \csname LTb\endcsname%
      \put(1694,484){\makebox(0,0){\strut{}$3.2$}}%
      \csname LTb\endcsname%
      \put(2068,484){\makebox(0,0){\strut{}$3.4$}}%
      \csname LTb\endcsname%
      \put(2442,484){\makebox(0,0){\strut{}$3.6$}}%
      \csname LTb\endcsname%
      \put(2958,484){\makebox(0,0){\strut{}$U_c^f$}}%
      \csname LTb\endcsname%
      \put(3190,484){\makebox(0,0){\strut{}$4$}}%
      \csname LTb\endcsname%
      \put(3564,484){\makebox(0,0){\strut{}$4.2$}}%
      \csname LTb\endcsname%
      \put(3938,484){\makebox(0,0){\strut{}$4.4$}}%
      \csname LTb\endcsname%
      \put(4312,484){\makebox(0,0){\strut{}$4.6$}}%
    }%
    \gplgaddtomacro\gplfronttext{%
      \csname LTb\endcsname%
      \put(198,2127){\rotatebox{-270}{\makebox(0,0){\strut{}$\beta^\zeta\Delta_0/U$}}}%
      \put(2722,154){\makebox(0,0){\strut{}$U$}}%
      \csname LTb\endcsname%
      \put(1870,3378){\makebox(0,0)[r]{\strut{}$\beta = 6$}}%
      \csname LTb\endcsname%
      \put(1870,3158){\makebox(0,0)[r]{\strut{}$\beta = 8$}}%
      \csname LTb\endcsname%
      \put(1870,2938){\makebox(0,0)[r]{\strut{}$\beta = 10$}}%
      \csname LTb\endcsname%
      \put(1870,2718){\makebox(0,0)[r]{\strut{}$\beta = 12$}}%
    }%
    \gplbacktext
    \put(0,0){\includegraphics{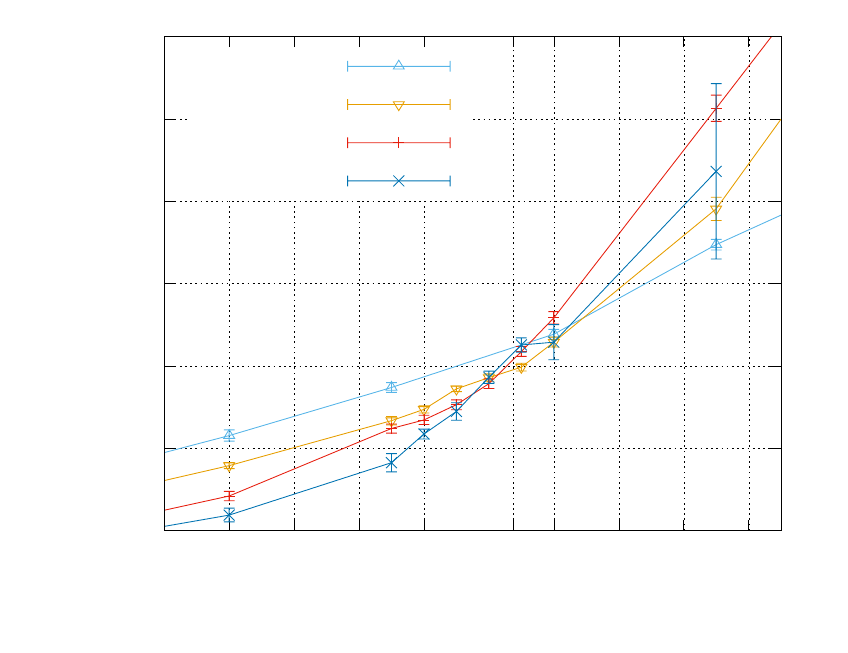}}%
    \gplfronttext
  \end{picture}%
\endgroup
 	\caption{Single-particle gap $\Delta_0$ scaled by $\beta$ (left panel) and by $\beta^\zeta/U$ (right panel), as a function of $U$. Note that $\Delta_0$ has been extrapolated to infinite volume
	and $\delta \to 0$ (continuum limit). Each line represents a given inverse temperature $\beta$. Here, $\zeta=\critZeta$ as obtained from the data collapse fit.
	The averages of all the crossing points $U_c^f$ and $U_c^g$ have been marked by vertical lines. For $U_c^g$, data with $\beta\le 4$ have been omitted due to the thermal gap discussed 
	in \Appref{thermal_gap}. All quantities are given in units of $\kappa$.
	\label{fig:crossing}}
\end{figure}

From the definitions of the scaling functions $F$ and $G$, one finds that they become independent of $\beta$ at $U = U_c$. Therefore, all lines in a 
$U$-$f$ and $U$-$g$ plot should cross at $U=U_c$. The average of all the crossing points (between pairs of lines with different $\beta$) yields an estimate of $U_c$ as well.
We obtain $U_c^{f}/\kappa=\critUf$ from the crossings in the $U$-$f$ plot (see \Figref{crossing}, left panel) and $U_c^{g}/\kappa=\critUg$ from the $U$-$g$ plot 
(see \Figref{crossing}, right panel). The errors are estimated from the standard deviation of all the crossings $\sigma_c$ and the number of crossings $n_c$ as $\sigma_c/\sqrt{n_c}$.
We find that all three results for $U_c$ are compatible within errors. However, the result from the data collapse analysis is much more precise than the crossing analysis because the data collapse analysis performs a global fit over all the MC data points, not just the points in the immediate vicinity of $U_c$.

\subsection{Zero temperature extrapolation}\label{sec:zero temperature extrapolation}

Let us first consider the scaling properties as a function of $\beta$ for $U = U_c$. As the scaling function reduces to a constant, we have
\begin{align}
\Delta_0^{} \propto \beta^{-1}\,.
\label{eq:scaling_beta}
\end{align}
For $U > U_c$ and $\beta \to \infty$, we have $\mathfrak{O} \sim (U-U_c)^q$ by construction
and therefore
\begin{align}
\Delta_0^{} & = \begin{cases}
0 & U \le U_c \\
c_U^{} (U-U_c)^{z\nu} & U > U_c
\end{cases}\label{eq:scaling_U} %
\end{align}
where $c_U^{}$ %
is a constant of proportionality.

We are now in a position to perform a simple extrapolation of $\Delta_0$ to $\beta \to \infty$, in order to visualize the quantum phase transition.
Let us assume that the scaling in Eq.~\eqref{eq:scaling_beta} holds approximately for $U$ slightly above $U_c$, up to a constant shift $\Delta_0^\infty$, 
the zero-temperature gap. Thus we fit
\begin{align}
\Delta_0^{} & = \Delta_0^\infty + c_\beta^{}\beta^{-1},
\end{align}
at a chosen value of $U/\kappa = 4$. The resulting value of $\Delta_0^\infty/\kappa=\DeltaInf$ then allows us to determine the 
coefficient $c_U = \coeffU$ (in units of $\kappa$) of Eq.~\eqref{eq:scaling_U}. This allows us to plot the 
solid black line with error bands in \Figref{delta_against_U}. Such an extrapolation should be regarded as valid only in the immediate 
vicinity of the phase transition. For $U \gg U_c$ the data seem to approach the mean-field result $\tilde\beta = 1/2$~\cite{Sorella_1992}. Furthermore, 
we note that an inflection point has been observed in~\Ref{Assaad:2013xua} at $U/\kappa\approx 4.1$, though this effect is thought to be an artifact of the extrapolation of the
MC data in the Trotter error $\delta$. The error estimation for the extrapolation $\beta \to \infty$ follows the same scheme as the one for the $u$-$g$ data collapse. 
We obtain the error band in \Figref{delta_against_U} as the area enclosed by the two lines corresponding to the lower bound of $U_c$ and the upper bound 
of $z\nu$ (on the left) and vice versa (on the right). This method conservatively captures any correlation between the different parameters. 

\section{Conclusions}\label{sec:conclusion}

Our work represents the first instance where the grand canonical BRS algorithm has been applied to the hexagonal Hubbard model 
(beyond mere proofs of principle), and we have found highly promising results. We emphasize that previously encountered issues related to the computational scaling and ergodicity 
of the HMC updates have been solved~\cite{Wynen:2018ryx}. 
We have primarily investigated the single-particle gap $\Delta$ (which we assume to be due to the semimetal-AFMI transition) as a function of $U/\kappa$, 
along with a comprehensive analysis of the temporal continuum, thermodynamic and zero-temperature limits. The favorable scaling of the HMC
enabled us to simulate lattices with $L > 100$ and to
perform a highly systematic treatment of all three limits. The latter limit was taken by means of a finite-size scaling analysis, which determines the
critical coupling $U_c/\kappa=\critU$ and the critical exponent $z\nu=\critNu$.
While we have not yet performed a direct MC calculation of the AFMI order parameter $m_s$, our scaling analysis of $\Delta/U$ has enabled an estimate of
the critical exponent $\tilde\beta=\critExp$. Depending on which symmetry is broken, the critical exponents of the hexagonal Hubbard model 
are expected to fall into one of the Gross-Neveu (GN) universality classes~\cite{PhysRevLett.97.146401}. The semimetal-AFMI transition should fall into the GN-Heisenberg 
$SU(2)$ universality class, as $m_s$ is described by a vector with three components. 

The GN-Heisenberg critical exponents have been studied by 
means of PMC simulations of the hexagonal Hubbard model, by the $d = 4-\epsilon$ expansion around the upper 
critical dimension $d$, by large~$N$ calculations, and by functional renormalization group (FRG) methods.
In Table~\ref{tab:crit_values}, we give an up-to-date comparison with our results. Our value for $U_c/\kappa$ is in overall agreement with previous MC simulations. 
For the critical exponents $\nu$ and $\tilde\beta$, the situation is less clear. Our results for $\nu$ (assuming $z = 1$ due to Lorentz invariance~\cite{PhysRevLett.97.146401}) and $\tilde\beta$ agree 
best with the HMC calculation (in the BSS formulation) of Ref.~\cite{Buividovich:2018yar}, followed by the FRG and large~$N$ calculations. 
On the other hand, our critical exponents are systematically larger than most PMC calculations and first-order $4-\epsilon$ expansion results. The agreement appears to be significantly 
improved when the $4-\epsilon$ expansion is taken to higher orders, although the discrepancy between expansions for $\nu$ and $1/\nu$ persists.
\begin{table}
	\caption{Summary of critical couplings $U_c/\kappa$ and critical exponents $\nu$ and $\beta$ (called $\tilde\beta$ in the rest of this work) obtained by recent MC calculations of 
	various Hubbard models in the Gross-Neveu (GN) Heisenberg universality class, and with other methods for direct calculations of the GN Heisenberg model.
	We include brief comments of special features of each
	calculation. Note the abbreviations HMC (Hybrid Monte Carlo), AF (Auxiliary Field), 
	BSS (Blankenbecler-Sugar-Scalapino) and BRS (Brower-Rebbi-Schaich). These concepts are explained in the main text. 
	Furthermore, we denote FRG (Functional Renormalization Group).
	Our value of $\nu$~($\dag$) is given for $z = 1$~\cite{PhysRevLett.97.146401}.
	Our estimate of $\beta$ is based on the mean-field result $m_s \sim \Delta/U$~($\ddag$)~\cite{Assaad:2013xua}.
	The asterisk (*) indicates that the $4-\epsilon$ exponents of Ref.~\cite{Herbut:2009vu} were used as input in the
	MC calculation of $U_c$ in Ref.~\cite{Assaad:2013xua}. Also, note the ambiguities~\cite{Otsuka:2015iba} as to the correct number of fermion components in the $4-\epsilon$ expansion
	of Ref.~\cite{Rosenstein:1993zf}. \label{tab:crit_values}}
	\vspace{.2cm}
	\begin{tabular}{lSSS}
		Method & $U_c/\kappa$ & $\nu$ & $\beta$ \\
		\hline
		Grand canonical BRS HMC (present work) & 3.834(14) & 1.185(43)$^\dag$ & 1.095(37)$^\ddag$ \\
		Grand canonical BSS HMC, complex AF~\cite{Buividovich:2018yar} & 3.90(5) & 1.162 & 1.08(2) \\
		Grand canonical BSS QMC~\cite{Buividovich:2018crq} & 3.94 & 0.93 & 0.75 \\
		Projection BSS QMC~\cite{Otsuka:2015iba} & 3.85(2) & 1.02(1) & 0.76(2) \\
		Projection BSS QMC, $d$-wave pairing field~\cite{Otsuka:2020lhc} & & 1.05(5) & \\
		Projection BSS QMC~\cite{Toldin:2014sxa} & 3.80(1) & 0.84(4) & 0.71(8) \\
		Projection BSS QMC, spin-Hall transition~\cite{Liu:2018sww} & & 0.88(7) & \\
		Projection BSS QMC, pinning field~\cite{Assaad:2013xua} & 3.78 & 0.882* & 0.794* \\
		GN $4-\epsilon$ expansion, 1st order~\cite{Herbut:2009vu, Otsuka:2015iba} & & 0.882* & 0.794* \\
		GN $4-\epsilon$ expansion, 1st order~\cite{Rosenstein:1993zf, Otsuka:2015iba} & & 0.851 & 0.824 \\
		GN $4-\epsilon$ expansion, 2nd order~\cite{Rosenstein:1993zf, Otsuka:2015iba} & & 1.01 & 0.995 \\
		GN $4-\epsilon$ expansion, $\nu$ 2nd order~\cite{Rosenstein:1993zf,heisenberg_gross_neveu} & & 1.08 & 1.06 \\
		GN $4-\epsilon$ expansion, $1/\nu$ 2nd order~\cite{Rosenstein:1993zf,heisenberg_gross_neveu} & & 1.20 & 1.17 \\
		GN $4-\epsilon$ expansion, $\nu$ 4th order~\cite{Zerf:2017zqi} & & 1.2352 & \\
		GN $4-\epsilon$ expansion, $1/\nu$ 4th order~\cite{Zerf:2017zqi} & & 1.5511 & \\
		GN FRG~\cite{heisenberg_gross_neveu} & & 1.31 & 1.32 \\
		GN FRG~\cite{Knorr:2017yze} & & 1.26 & \\
		GN Large $N$~\cite{Gracey:2018qba} & & 1.1823 &
	\end{tabular}
\end{table}

Our results show that the BRS algorithm is now applicable to problems of a realistic size in the field of carbon-based nano-materials.
There are several future directions in which our present work can be developed. For instance, while the AFMI phase may not be directly observable in graphene, 
we note that tentative empirical evidence for such a phase exists in carbon nanotubes~\cite{Deshpande02012009}, along with preliminary theoretical 
evidence from MC simulations presented in Ref.~\cite{Luu:2015gpl}. The MC calculation of the single-particle Mott gap in a (metallic) carbon nanotube is 
expected to be much easier, since the lattice dimension $L$ is determined by the physical nanotube radius used in the experiment (and by the number of unit 
cells in the longitudinal direction of the tube). As electron-electron interaction (or correlation) effects are expected to be more pronounced 
in the (1-dimensional) nanotubes, the treatment of flat graphene as the limiting case of an infinite-radius nanotube would be especially interesting.
Strong correlation effects could be even more pronounced in the (0-dimensional) carbon fullerenes (buckyballs), where we are also faced 
with a fermion sign problem due to the admixture of pentagons into the otherwise-bipartite honeycomb structure~\cite{Wynen:2020uzx}.
This sign problem has the unusual property of vanishing as the system size becomes
large, as the number of pentagons in a buckyball is fixed by its Euler characteristic to be exactly~$12$, independent of the number of hexagons.
The mild scaling of HMC with system size gives access to very large physical systems ($\sim 10^4$ sites or more), so it may be plausible to put a 
particular experimental system (a nanotube, a graphene patch, or a topological insulator, for instance) into software, for a direct, first-principles 
Hubbard model calculation.

\section*{Acknowledgements}

We thank Jan-Lukas Wynen for helpful discussions on the Hubbard model and software issues. We also thank Michael Kajan for proof reading and for providing a lot of detailed comments.
This work was funded, in part, through financial support from the Deutsche Forschungsgemeinschaft (Sino-German CRC 110 and SFB TRR-55).
E.B. is supported by the U.S. Department of Energy under Contract No. DE-FG02-93ER-40762.
The authors gratefully acknowledge the computing time granted through JARA-HPC on the supercomputer JURECA~\cite{jureca} at Forschungszentrum J\"ulich.
We also gratefully acknowledge time on DEEP~\cite{DEEP}, an experimental modular supercomputer at the J\"ulich Supercomputing Centre.

\appendix

\section{Subtleties of the mixed differencing scheme 
\label{sec:mixed_bias}}

As in Ref.~\cite{Luu:2015gpl}, we have used a mixed-differencing scheme for the $A$ and $B$ sublattices, which was first suggested by
Brower \textit{et al.} in Ref.~\cite{Brower:2012zd}. While mixed differencing does not cancel the linear Trotter error completely, it does diminish it significantly,
as shown in Ref.~\cite{Luu:2015gpl}. We shall now discuss some fine points related to the correct continuum limit when mixed differencing is used.

\subsection{Seagull term}

Let us consider the forward differencing used in Ref.~\cite{Luu:2015gpl} for the $AA$ contribution to fermion operator,
\begin{equation}
M^{AA}_{(x,t)(y,t^\prime)} = \delta_{xy}^{} \left\{
-\delta_{t,t^\prime}^{} + \left[\exp(i\phidelta_{x,t}^{})-\mdelta\right] \delta_{t+1,t^\prime}^{} \right\},
\label{ferm_op}
\end{equation}
with the gauge links and an explicit staggered mass $\mdelta$ (not to be confused with the AFMI order parameter) on the time-off-diagonal. We recall
that a tilde means that the corresponding quantity is multiplied by $\delta=\beta/N_t$. An expansion in $\delta$ gives
\begin{align}
M^{AA}_{(x,t)(y,t^\prime)} 
& = \delta_{xy}^{} \left\{
-\delta_{t,t^\prime} + \left[1 + i\phidelta_{x,t}^{} - \frac{\phidelta_{x,t}^2}{2} - \mdelta\right] \delta_{t+1,t^\prime}^{} \right\} + \ordnung{\delta^3}, \\
& = \delta_{xy}^{} \left\{
\delta_{t+1,t^\prime} - \delta_{t,t^\prime} + \left[ i\phidelta_{x,t}^{} - \frac{\phidelta_{x,t}^2}{2} - \mdelta\right] \delta_{t+1,t^\prime}^{} \right\} + \ordnung{\delta^3}, \\
& = \delta_{xy}^{} \left\{
\delta\partial_t^{} +\frac{\delta^2\partial^2_t}{2} + \left[ i\phidelta_{x,t}^{} - \frac{\phidelta_{x,t}^2}{2} - \mdelta\right] 
(\delta_{t,t^\prime}^{} + \delta\partial_t^{})\right\} + \ordnung{\delta^3}, \\
& = \delta_{xy}^{} \left\{
\delta\partial_t^{} + ( i\phidelta_{x,t}^{} - \mdelta^\prime ) \delta_{t,t^\prime} - \delta\mdelta\partial_t^{} \right\} + \ordnung{\delta^3},
\end{align}
where terms proportional to $\delta$ remain the continuum limit. In the last step we defined
\begin{equation}
\mdelta^\prime := \mdelta + \frac{\phidelta_{x,t}^2}{2} - \frac{\delta^2\partial^2_t}{2} - i\delta\phidelta_{x,t}^{}\partial_t^{},
\label{eqn_effective_staggered_mass}
\end{equation}
as the ``effective'' staggered mass at finite $\delta$.
The same calculation can be performed for the $BB$ contribution of Ref.~\cite{Luu:2015gpl}. This gives
\begin{align}
M^{BB}_{(x,t)(y,t^\prime)}
& = \delta_{xy}^{} \left\{
\delta\partial_t^{} + ( i\phidelta_{x,t}^{} + \mdelta^\prime ) \delta_{t,t^\prime} - \delta\mdelta\partial_t^{} \right\} + \ordnung{\delta^3},
\end{align}
where the effective staggered mass has the opposite sign, as expected.

Let us discuss the behavior of $\mdelta^\prime$ when $m_s \to 0$ and $\delta \to 0$. First, 
$\partial_t^2$ is negative semi-definite, $\phidelta_{x,t}^2$ is positive semi-definite, and $i\phidelta_{x,t}\partial_t$ is indefinite, 
but one typically finds that $\mdelta^\prime \ge \mdelta$. For a vanishing bare staggered mass $\mdelta \to 0$, this creates a non-vanishing 
bias between the sublattices at $\delta \neq 0$, which is due to the mixed differencing scheme. Numerically, we find that this effect prefers $\erwartung{m_A-m_B}>0$.
Second, the ``seagull term'' $\phidelta_{x,t}^2$ is not suppressed in the continuum limit, as first noted in Ref.~\cite{Brower:2012zd}. This happens because,
in the vicinity of continuum limit, the Gaussian part of the action becomes narrow, and 
$\phidelta$ is approximately distributed as $\phidelta \sim \mathcal{N}(0,\sqrt{\delta U})$. Because 
$\phidelta$ scales as $\sqrt{\delta}$, the seagull term is not in $\ordnung{\delta^2}$ but effectively linear in $\delta$, denoted as $\efford{\delta}$.
The seagull term contains important physics, and should be correctly generated by the gauge links.

\subsection{Field redefinition
\label{sec_field_redef}}

Following Brower \textit{et al.} in \Ref{Brower:2012zd}, the seagull term can be absorbed by means of a redefinition of the Hubbard-Stratonovich field,
at the price of generating the so-called ``normal-ordering term'' of Ref.~\cite{Brower:2012zd}, which is of
physical significance. Let us briefly consider how this works in our case. The field redefinition of \Ref{Brower:2012zd} is
\begin{equation}
\phi_x^{} := \varphi_x^{} - \frac{\delta}{2} V_{xy}^{} \varphi_y^{} \psi^*_y \psi_y^{},
\label{phi_shift}
\end{equation}
in terms of the field $\psi$ on which the fermion matrix $M$ acts. For backward differencing of $M$, the full Hamiltonian
includes the terms
\begin{align}
& \frac{1}{2} \sum_t \tilde\phi_{x,t}^{} \tilde V_{xy}^{-1} \tilde \phi_{y,t}^{} +
\sum_{t,t^\prime} \psi^*_{x,t} \bigg[ \delta_{t,t^\prime}^{}
-\exp(-i\tilde\phi_{x,t}^{}) \delta_{t-1,t^\prime}^{} + \tilde m \delta_{t-1,t^\prime}^{} \bigg] \psi_{x,t^\prime}^{}
\nonumber \\
& \qquad =
\ { \frac{1}{2} \int dt \, \phi_x^{} V_{xy}^{-1} \phi_y^{}}
\ { + \int dt \, \psi^*_x \partial_t^{} \psi_x^{}}
\ { + \, i \int dt \, \psi^*_x \phi_x^{} \psi_x^{}}
\ { + \, m \int dt \, \psi^*_x \psi_x^{}}
\ { + \,\frac{\delta}{2} \int dt \, \psi^*_x \phi_x^2 \psi_x^{}}
+ \efford{\delta},
\label{lattice_B}
\end{align}
where the Gaussian term generated by the Hubbard-Stratonovich transformation is included, and $\tilde m$ is left unspecified for the moment.
If we apply the redefinition~\eqref{phi_shift} to the Gaussian term, we find
\begin{align}
\frac{1}{2} \int dt \, \phi_x^{} V_{xy}^{-1} \phi_y^{} & = \frac{1}{2} \int dt \, \varphi_x^{} V_{xy}^{-1} \varphi_y^{}
- \frac{\delta}{2} \int dt \, \varphi_x^{} V_{xz}^{-1} V_{zy}^{} \varphi_y^{} \psi^*_y \psi_y^{} + \efford{\delta},
\nonumber \\
& = \frac{1}{2} \int dt \, \varphi_x^{} V_{xy}^{-1} \varphi_y^{}
- \frac{\delta}{2} \int dt \, \psi^*_x \varphi_x^2 \psi_x^{} + \efford{\delta},
\end{align}
and along the lines of \Ref{Brower:2012zd}, we note that the $\varphi_x^2$ seagull term cancels the $\phi_x^2$ 
seagull term of Eq.~(\ref{lattice_B}) to leading order in $\delta$. As we are performing a path integral over the Hubbard-Stratonovich field,
we need to account for the Jacobian of the field redefinition, which is
\begin{align}
\int \mathcal{D}\tilde\phi =
\int \mathcal{D}\tilde \varphi \, \det \left[\frac{\partial\tilde\phi_{x,t}^{}}{\partial\tilde\varphi_{y,t^\prime}^{}}\right]
& = \int \mathcal{D}\tilde\varphi \, \exp \left[\mathrm{Tr} \log\left(\delta_{xy}^{}\delta_{t,t^\prime}^{} - \frac{\delta}{2}
V_{xy}^{} \psi^*_{y,t} \psi_{y,t}^{} \delta_{t,t^\prime}^{}\right) \right]
\nonumber \\
& \simeq \int \mathcal{D}\varphi \,
\exp\left(- \frac{1}{2} \int dt \,  V_{xx}^{} \psi^*_x \psi_x^{}\right),
\end{align}
where in the last step we used $\log(1+\delta z) = \delta z+\ordnung{\delta^2}$ before taking the continuum limit. 
We conclude that the seagull term in the expansion of the gauge links has the correspondence
\begin{equation}
\frac{\delta}{2} \int dt \, \psi^*_x \phi_x^2 \psi_x^{}
\longleftrightarrow \frac{1}{2} \int dt \, V_{xx}^{} \psi^*_x \psi_x^{},
\label{seagull_norm}
\end{equation}
which is exactly the normal-ordering term proportional to $V_{xx}^{}/2$ of \Ref{Smith:2014tha}.
Hence, as argued in \Ref{Brower:2012zd}, the normal-ordering term should be omitted when gauge links are used, as an equivalent term is dynamically
generated by the gauge links.
This statement is valid when backward differencing is used for both sublattices. In \Appref{alternative derivative}, we discuss
how this argument carries over to the case of forward and mixed differencing.

\subsection{Alternative forward difference 
\label{sec:alternative derivative}}

In case the backward differencing of Eq.~(\ref{lattice_B}) is used for both sublattices as in \Ref{Smith:2014tha}, then simply taking
the usual staggered mass term
\begin{equation}
\tilde m = \mdelta \quad (x \in A),
\qquad
\tilde m = -\mdelta \quad (x \in B),
\end{equation}
suffices to get the correct Hubbard Hamiltonian, as both sublattices receive a dynamically generated normal-ordering term with coefficient $V_{xx}^{}/2$.
However, the mixed-difference lattice action in \Ref{Luu:2015gpl} produces a ``staggered'' normal-ordering term, with $-V_{xx}^{}/2$ for sublattice $A$ and 
$V_{xx}^{}/2$ for sublattice $B$. Hence, with the mixed-difference operator of \Ref{Luu:2015gpl} (forward for sublattice $A$, backward for sublattice $B$), 
we should instead take
\begin{equation}
\tilde m = \tilde V_{00}^{} + \mdelta \quad (x \in A),
\qquad
\tilde m = -\mdelta \quad (x \in B),
\end{equation}
in order to again obtain the physical Hubbard Hamiltonian with normal-ordering and staggered mass terms.
Therefore, in our current work we adopt the alternative forward differencing
\begin{align}
M^{AA}_{(x,t)(y,t^\prime)} & = \delta_{xy}^{} \left\{ \delta_{t+1,t^\prime}^{}
- \left[\exp(-i\phidelta_{x,t}^{}) + \mdelta\right] \delta_{t,t^\prime}^{} \right\},
\label{MAA_modified}
\end{align}
instead of Eq.~\eqref{ferm_op}, which again yields a normal-ordering term $V_{xx}^{}/2$ for sublattice $A$. As in \Ref{Smith:2014tha}, we thus
retain the desirable feature of a completely dynamically generated normal-ordering term. In our actual numerical simulations, we set the bare 
staggered mass $\mdelta = 0$. In our CG solver with Hasenbusch preconditioning, we work with finite $\mdelta$~\cite{hasenbusch}.
The spectrum of the operator~\eqref{MAA_modified} lacks conjugate reciprocity, which causes an ergodicity problem~\cite{Wynen:2018ryx}.
\section{Finding a plateau 
\label{sec:plateau_estimation}}

Here we present an automatized, deterministic method that reliably finds the optimal plateau in a given data set (such as the effective mass $m(\tau)$). Specifically, our method finds 
the region of least slope and fluctuations, and checks whether this region is a genuine plateau without significant drift. If a given time series does not exhibit an acceptable plateau, 
our method returns an explicit error message.

Apart from the time series $m(\tau)$ expected to exhibit a plateau, the algorithm requires two parameters to be chosen in advance. 
The first is the minimal length $\lambda$ a plateau should have. The second is an ``analysis window'' of width $\mu\le\lambda$. 
This controls how many data points are considered in the analysis of local fluctuations. We find that
\begin{align}
\lambda & = \frac{N_t^{}}{6}, \\
\mu & = \log_2^{}(N_t^{}),
\end{align}
are in most cases good choices.

Algorithm~\ref{alg_plateau} describes the procedure in detail. The idea is to find a balance between least statistical and systematic fluctuations. 
Statistical fluctuations decrease with increasing plateau length. This is why we seek to choose the plateau as long as possible, without running into a region 
with large systematic deviations. This property can also be used to our advantage. If we calculate the mean from a given time $\tau_2$ to all the 
previous times, the influence of another point compatible with the mean will decrease with the distance from $\tau_2$. Thus the local fluctuation of the 
running mean decreases, until it reaches a point with significant systematic deviation. This local fluctuation minimum marks the optimal $\tau_1$. The plateau 
then ranges from $\tau_1$ to $\tau_2$. We check that it does not exhibit significant drift, by fitting a linear function and checking if the first order term deviates 
from zero within twice its error. By repeating the analysis for all possible values of $\tau_2$, the globally best plateau can be found, as determined by 
least local fluctuations of the running mean.

	\begin{algorithm}
		\SetKwInOut{Input}{input}\SetKwInOut{Output}{output}
		\Input{$N_t$, $m[0,\dots,\,N_t-1]$, $\lambda$, $\mu$}
		\Output{$\tau_1$, $\tau_2$}
		\BlankLine
		\For{$\tau'=\mu-1,\dots,\,N_t-1$}{
			\For{$\tau=0,\dots,\, \tau'$}{
				$\overline{m}[\tau;\tau'] = \mathrm{mean}\!\left(m[\tau,\dots,\tau']\right)$\;
			}
			\For{$\tau=0,\dots,\, \tau'-\mu+1$}{
				$\sigma[\tau;\tau'] = \mathrm{sd}\!\left(\overline{m}[\tau,\dots,\tau+\mu-1;\tau']\right)$\;
			}
			$\tau_1^*[\tau'] = \underset{\tau\in \left\{0,\dots\,, \tau'-\mu\right\}}{\mathrm{argmin}} \left(\sigma[\tau;\tau']\right)$\;
		}
		$\Lambda_{\phantom{0}} = \left\{\left(\tau,\tau'\right)\,|\:\tau=\tau_1^*[\tau'],\,\tau'-\tau\ge\lambda\right\}$\;
		$\Lambda_0 = \left\{\left(\tau,\tau'\right)\in \Lambda\,|\:m[\tau,\dots,\tau'] \text{ has no significant drift}\right\}$\;
		\uIf{$\Lambda_0 \neq \varnothing$}{
			$\left(\tau_1,\tau_2\right) = \underset{(\tau,\tau')\in \Lambda_0}{\mathrm{argmin}} \left(\sigma[\tau;\tau']\right)$\;
		}\Else{
			No acceptable plateau of requested length found.
		}
		\caption{Finding a fit range for a plateau in a time series. \label{alg_plateau}}
	\end{algorithm}

As every range in the set $\Lambda_0$ from Algorithm~\ref{alg_plateau} is a valid plateau, it allows us to estimate the systematic error due to the choice of plateau. 
We simply repeat the calculation of the relevant observable for all ranges in $\Lambda_0$, and interpret the standard deviation of the resulting set of values as a systematic uncertainty.
\section{Thermal gap}\label{sec:thermal_gap}

It is useful to consider the influence of the inverse temperature $\beta$ on the single-particle gap, in order to provide a better understanding
of the scaling of $\Delta$ with $\beta$. Naturally, we are not able to solve the entire problem analytically, so we shall consider small
perturbations in the coupling $U$, and assume that the dispersion relation of graphene is not significantly perturbed by the interaction (which is
expected to be the case when $U$ is small). Let us now compute the expectation value of the number of electrons excited from the ground state.
As we consider exclusively the conduction band, we assume that the particle density follows Fermi-Dirac statistics.  We take the positive-energy part of
\begin{equation}
\omega_k^{} := \kappa\sqrt{\tilde\omega_k^2},
\qquad
\tilde\omega_k^2 = 
3 + 4 \cos(3a k_x^{}/2) \cos(\sqrt{3}a k_y^{}/2) + 2\cos(\sqrt{3} a k_y^{}),
\end{equation}
where $a \simeq 1.42$~{\AA} is the nearest-neighbor lattice spacing and we assume that every excited electron contributes an energy $E(U)$ to a ``thermal gap'' $\Delta(\beta)$.
These considerations yield the gap equation
\begin{equation}
\Delta(\beta) = E(U) a^2 f_{\text{BZ}}^{} \int_{k\in\text{BZ}}
\frac{d^2k}{(2\pi)^2} \frac{1}{1+\exp(\beta\omega_k^{})},
\qquad
f_{\text{BZ}}^{} := \frac{3\sqrt{3}}{2},
\label{eqn_therm_gap_inf_L}
\end{equation}
where the factor $f_{\text{BZ}}^{}$ is due to the 
hexagonal geometry of the first Brillouin zone (BZ). It should be emphasized that the thermal gap is not the interaction-driven Mott gap we are studying here, 
even though it is not numerically distinguishable from the latter. A thermal gap can occur even if the conduction and valence bands touch or overlap. 
The physical interpretation of the thermal gap (as explained above) is a measure of the degree of 
excitation above the ground state, based on the number of excited states that are already occupied in thermal equilibrium.

\subsection{Finite temperature}

\begin{figure}[t]
\begingroup
  \inputencoding{latin1}%
  \makeatletter
  \providecommand\color[2][]{%
    \GenericError{(gnuplot) \space\space\space\@spaces}{%
      Package color not loaded in conjunction with
      terminal option `colourtext'%
    }{See the gnuplot documentation for explanation.%
    }{Either use 'blacktext' in gnuplot or load the package
      color.sty in LaTeX.}%
    \renewcommand\color[2][]{}%
  }%
  \providecommand\includegraphics[2][]{%
    \GenericError{(gnuplot) \space\space\space\@spaces}{%
      Package graphicx or graphics not loaded%
    }{See the gnuplot documentation for explanation.%
    }{The gnuplot epslatex terminal needs graphicx.sty or graphics.sty.}%
    \renewcommand\includegraphics[2][]{}%
  }%
  \providecommand\rotatebox[2]{#2}%
  \@ifundefined{ifGPcolor}{%
    \newif\ifGPcolor
    \GPcolortrue
  }{}%
  \@ifundefined{ifGPblacktext}{%
    \newif\ifGPblacktext
    \GPblacktexttrue
  }{}%
  \let\gplgaddtomacro\g@addto@macro
  \gdef\gplbacktext{}%
  \gdef\gplfronttext{}%
  \makeatother
  \ifGPblacktext
    \def\colorrgb#1{}%
    \def\colorgray#1{}%
  \else
    \ifGPcolor
      \def\colorrgb#1{\color[rgb]{#1}}%
      \def\colorgray#1{\color[gray]{#1}}%
      \expandafter\def\csname LTw\endcsname{\color{white}}%
      \expandafter\def\csname LTb\endcsname{\color{black}}%
      \expandafter\def\csname LTa\endcsname{\color{black}}%
      \expandafter\def\csname LT0\endcsname{\color[rgb]{1,0,0}}%
      \expandafter\def\csname LT1\endcsname{\color[rgb]{0,1,0}}%
      \expandafter\def\csname LT2\endcsname{\color[rgb]{0,0,1}}%
      \expandafter\def\csname LT3\endcsname{\color[rgb]{1,0,1}}%
      \expandafter\def\csname LT4\endcsname{\color[rgb]{0,1,1}}%
      \expandafter\def\csname LT5\endcsname{\color[rgb]{1,1,0}}%
      \expandafter\def\csname LT6\endcsname{\color[rgb]{0,0,0}}%
      \expandafter\def\csname LT7\endcsname{\color[rgb]{1,0.3,0}}%
      \expandafter\def\csname LT8\endcsname{\color[rgb]{0.5,0.5,0.5}}%
    \else
      \def\colorrgb#1{\color{black}}%
      \def\colorgray#1{\color[gray]{#1}}%
      \expandafter\def\csname LTw\endcsname{\color{white}}%
      \expandafter\def\csname LTb\endcsname{\color{black}}%
      \expandafter\def\csname LTa\endcsname{\color{black}}%
      \expandafter\def\csname LT0\endcsname{\color{black}}%
      \expandafter\def\csname LT1\endcsname{\color{black}}%
      \expandafter\def\csname LT2\endcsname{\color{black}}%
      \expandafter\def\csname LT3\endcsname{\color{black}}%
      \expandafter\def\csname LT4\endcsname{\color{black}}%
      \expandafter\def\csname LT5\endcsname{\color{black}}%
      \expandafter\def\csname LT6\endcsname{\color{black}}%
      \expandafter\def\csname LT7\endcsname{\color{black}}%
      \expandafter\def\csname LT8\endcsname{\color{black}}%
    \fi
  \fi
    \setlength{\unitlength}{0.0500bp}%
    \ifx\gptboxheight\undefined%
      \newlength{\gptboxheight}%
      \newlength{\gptboxwidth}%
      \newsavebox{\gptboxtext}%
    \fi%
    \setlength{\fboxrule}{0.5pt}%
    \setlength{\fboxsep}{1pt}%
\begin{picture}(7200.00,5040.00)%
    \gplgaddtomacro\gplbacktext{%
      \csname LTb\endcsname%
      \put(682,704){\makebox(0,0)[r]{\strut{}$0$}}%
      \csname LTb\endcsname%
      \put(682,1527){\makebox(0,0)[r]{\strut{}$5$}}%
      \csname LTb\endcsname%
      \put(682,2350){\makebox(0,0)[r]{\strut{}$10$}}%
      \csname LTb\endcsname%
      \put(682,3173){\makebox(0,0)[r]{\strut{}$15$}}%
      \csname LTb\endcsname%
      \put(682,3996){\makebox(0,0)[r]{\strut{}$20$}}%
      \csname LTb\endcsname%
      \put(682,4819){\makebox(0,0)[r]{\strut{}$25$}}%
      \csname LTb\endcsname%
      \put(814,484){\makebox(0,0){\strut{}$0$}}%
      \csname LTb\endcsname%
      \put(1623,484){\makebox(0,0){\strut{}$0.5$}}%
      \csname LTb\endcsname%
      \put(2433,484){\makebox(0,0){\strut{}$1$}}%
      \csname LTb\endcsname%
      \put(3242,484){\makebox(0,0){\strut{}$1.5$}}%
      \csname LTb\endcsname%
      \put(4051,484){\makebox(0,0){\strut{}$2$}}%
      \csname LTb\endcsname%
      \put(4861,484){\makebox(0,0){\strut{}$2.5$}}%
      \csname LTb\endcsname%
      \put(5670,484){\makebox(0,0){\strut{}$3$}}%
      \csname LTb\endcsname%
      \put(6479,484){\makebox(0,0){\strut{}$3.5$}}%
    }%
    \gplgaddtomacro\gplfronttext{%
      \csname LTb\endcsname%
      \put(198,2761){\rotatebox{-270}{\makebox(0,0){\strut{}$\beta^2\Delta_0$}}}%
      \put(3808,154){\makebox(0,0){\strut{}$U$}}%
      \csname LTb\endcsname%
      \put(3982,4646){\makebox(0,0)[r]{\strut{}$\beta = 3$}}%
      \csname LTb\endcsname%
      \put(3982,4426){\makebox(0,0)[r]{\strut{}$\beta = 4$}}%
      \csname LTb\endcsname%
      \put(3982,4206){\makebox(0,0)[r]{\strut{}$\beta = 6$}}%
      \csname LTb\endcsname%
      \put(3982,3986){\makebox(0,0)[r]{\strut{}$\beta = 8$}}%
      \csname LTb\endcsname%
      \put(3982,3766){\makebox(0,0)[r]{\strut{}$\beta = 10$}}%
      \csname LTb\endcsname%
      \put(3982,3546){\makebox(0,0)[r]{\strut{}$\beta = 12$}}%
      \csname LTb\endcsname%
      \put(3982,3326){\makebox(0,0)[r]{\strut{}thermal gap, E(U)=5.95U}}%
    }%
    \gplbacktext
    \put(0,0){\includegraphics{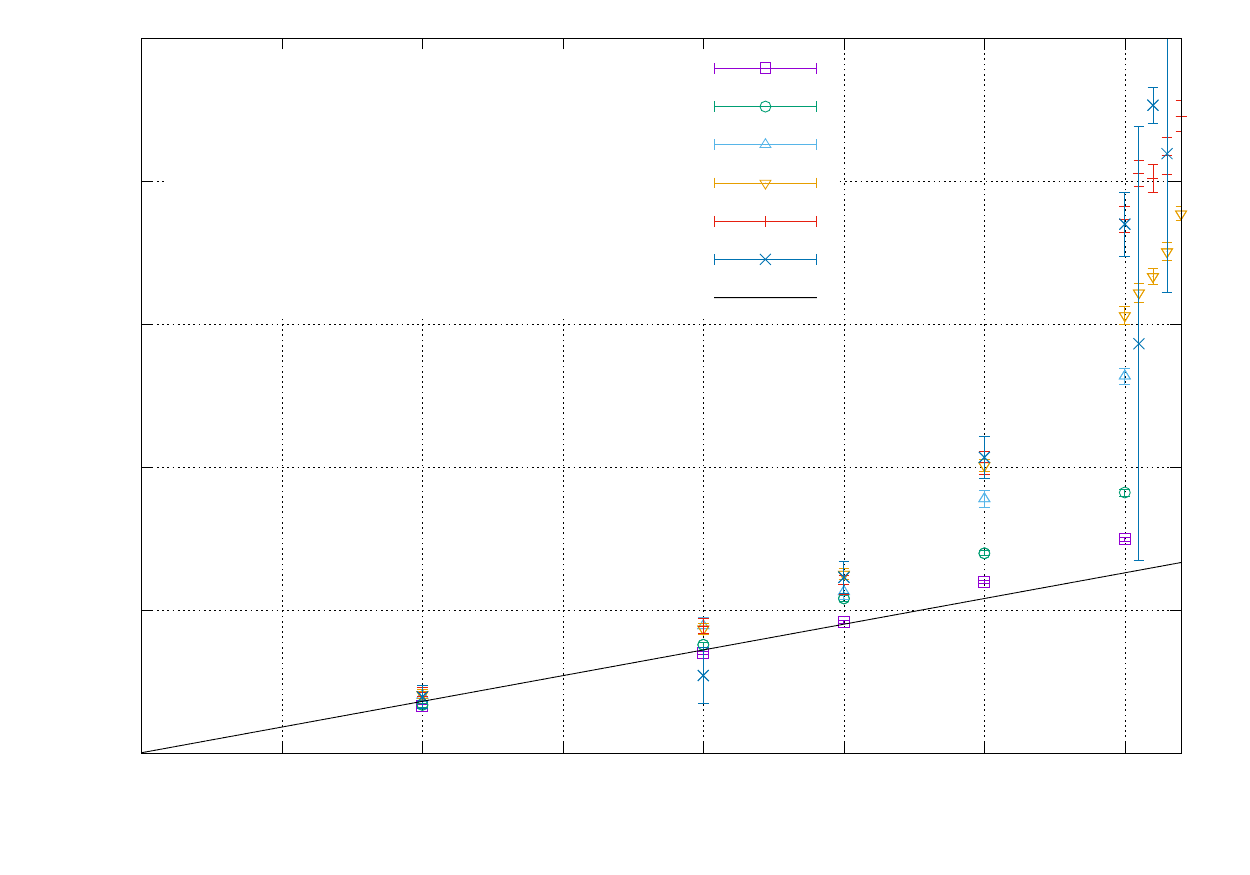}}%
    \gplfronttext
  \end{picture}%
\endgroup
 	\caption{Illustration of the thermal gap in the weakly coupled regime, as given by Eq.~\eqref{eqn:thermal_gap}. Our MC data for the single-particle gap $\Delta$ from
	\Figref{delta_against_U} is shown multiplied by $\beta^2$. All quantities are expressed in appropriate units of $\kappa$.}
	\label{fig:beta_sq_delta_against_U}
\end{figure}

Let us evaluate Eq.~(\ref{eqn_therm_gap_inf_L}) under the assumption that $\beta$ is large.
Then, the integrand only contributes in the region where $\omega_k^{} \approx 0$, in other words 
near the Dirac points $K$ and $K^\prime$ located at momenta $k_D^{}$.
In the vicinity of a Dirac point, the dispersion relation reduced to the well-known Dirac cone 
\begin{equation}
\omega_k^{} \simeq v_F^{} | k-k_D^{} |,
\qquad
v_F^{} := 3\kappa a/2,
\end{equation}
with Fermi velocity $v_F$. Within this approximation, we may 
sum over the two Dirac points and perform the angular integral, which gives
\begin{align}
\Delta(\beta) & \approx
2 E(U) a^2 f_{\text{BZ}}^{} \int_{k\in\mathbb{R}^2}
\frac{d^2k}{(2\pi)^2} 
\frac{1}{1+\exp(\beta v_F^{} | k | )}
= E(U) a^2 f_{\text{BZ}}^{}
\int_{0}^\infty \frac{dk}{\pi}
\frac{k}{1+\exp(\beta v_F^{} k)}, \\
\label{eqn:thermal_gap}
& = E(U) a^2 f_{\text{BZ}}^{} \times \frac{\pi}{12} (\beta v_F^{})^{-2}
= \frac{\sqrt{3}\pi}{18} \, E(U) (\beta\kappa)^{-2} 
\approx \num{0.3023} \, E(U) (\beta\kappa)^{-2},
\end{align}
where the Fermi-Dirac integral has been evaluated in terms of the polylogarithm function. We expect the
error of this approximation to be exponentially suppressed in $\beta$.

In \Figref{beta_sq_delta_against_U}, we validate Eq.~\eqref{eqn:thermal_gap} using our data for $\Delta$ shown in \Figref{delta_against_U}.
We find that the prediction of quadratic scaling in $\beta$ and the linear approximation in $U$ are quite accurate.
A fit of Eq.~\eqref{eqn:thermal_gap} to our MC data for $\Delta$ in the weak-coupling regime $U/\kappa \le 2$ gives
\begin{align}
E(U) = \num{5.95\pm.15}\,U,
\end{align}
under the (perturbative) assumption $E(U)\propto U$. This fit is shown in \Figref{beta_sq_delta_against_U}, where we plot $\beta^2\Delta$ as a function
of $U$ (in proper units of $\kappa$). In the weakly coupled regime, the MC data for $\beta^2\Delta$ coincide and fall on a straight line. Once the critical coupling $U_c$ is
approached, the points for various $\beta$ separate. As expected, the linear dependence on $U$ persists longest for small $\beta$, as temperature effects 
are dominant over interaction effects. The quadratic scaling with $\beta$ is most accurate for large $\beta$, which is in line with the expected
exponential convergence stated above.

\subsection{Finite lattice size}

Let us also consider the leading correction to the thermal gap due to finite lattice size $L$.
The discretized form of Eq.~(\ref{eqn_therm_gap_inf_L}) is
\begin{equation}
\Delta(L,\beta)=\frac{E(U)}{L^2}\sum_{k\in \text{BZ}}\frac{1}{1+\exp(\beta\omega_k^{})}
\label{eqn_therm_gap_discrete}
\end{equation}
which in general has a very complicated convergence behavior. 
Because of periodic boundary conditions, Eq.~(\ref{eqn_therm_gap_discrete}) is
an effective trapezoidal approximation to Eq.~(\ref{eqn_therm_gap_inf_L}), thus the
convergence is \textit{a priori} expected to scale as $\ordnung{L^{-2}}$.

We shall now obtain a precise leading-order error estimation. Let us discretize the 
first BZ is discretized into a regular triangular lattice, with lattice spacing $h\propto L^{-1}$. 
We integrate a function $f(x,y)$ over a single triangle, spanned by the coordinates $(\pm h/2, 0)$ 
and $(0, \sqrt{3}h/2)$,
\begin{equation}
I := \int_{-h/2}^{h/2} dx \int_0^{b(x)} dy \, f(x,y),
\qquad
b(x) := \sqrt{3}h/2-\sqrt{3} | x |,
\end{equation}
and subtract the average of $f(x,y)$ over the corner points multiplied by the area of the triangle,
\begin{equation}
\hat{I} := \frac{\sqrt{3}h^2}{4} \times \frac{1}{3}
\left[ f(-h/2,0)+f(h/2,0)+f(0,\sqrt{3}h/2) \right],
\end{equation}
which gives the (local) error
\begin{equation}
\delta I := 
I-\hat{I}=-\frac{\sqrt{3}}{64} \left[\delsqu{f}{x}(0)+\delsqu{f}{y}(0)\right] h^4 + \ordnung{h^5},
\end{equation}
due to discretization. The global error is obtained by summing over the complete BZ,
\begin{align}
\sum_{k\in \text{BZ}} \delta I(k) & =
-\frac{\sqrt{3}}{64}\sum_{k\in \text{BZ}} \left[\delsqu{f}{x}(k)+\delsqu{f}{y}(k)\right] h^4 + \ordnung{L^2h^5}, \\
&\propto \frac{1}{L^4} \sum_{k\in \text{BZ}} \left[\delsqu{f}{x}(k)+\delsqu{f}{y}(k)\right] + \ordnung{L^{-3}}, \\
&\propto \frac{1}{L^2} \int_{k\in \text{BZ}} d^2k \left[\delsqu{f}{x}(k)+\delsqu{f}{y}(k)\right] + \ordnung{L^{-3}},
\end{align}
which equals
\begin{align}
\sum_{k\in \text{BZ}} \delta I(k) & \propto
\frac{1}{L^2} \oint_{k\in \partial\text{BZ}} \nabla f(k) \cdot d\vec k + \ordnung{L^{-3}},
\label{eqn_integral_1st_bz} \\
& \propto \ordnung{L^{-3}},
\end{align}
where Gauss's theorem has been applied in Eq.~\eqref{eqn_integral_1st_bz}. Hence, 
the projection of $\nabla f$ onto the normal of the BZ is integrated over the boundary of the BZ.
As every momentum-periodic function takes the same values on the opposite edges of the BZ, the result sums up to zero.
Surprisingly, one then finds that the second order error term in $L$ vanishes. For the special case of
$f(k)\propto 1/(1+\exp(\beta\omega_k^{}))$, the gradient in BZ-normal direction vanishes everywhere on the boundary, and the integral in Eq.~\eqref{eqn_integral_1st_bz} 
is trivially zero.

\begin{figure}[t]
\centering
\includegraphics[width=.45\textwidth]{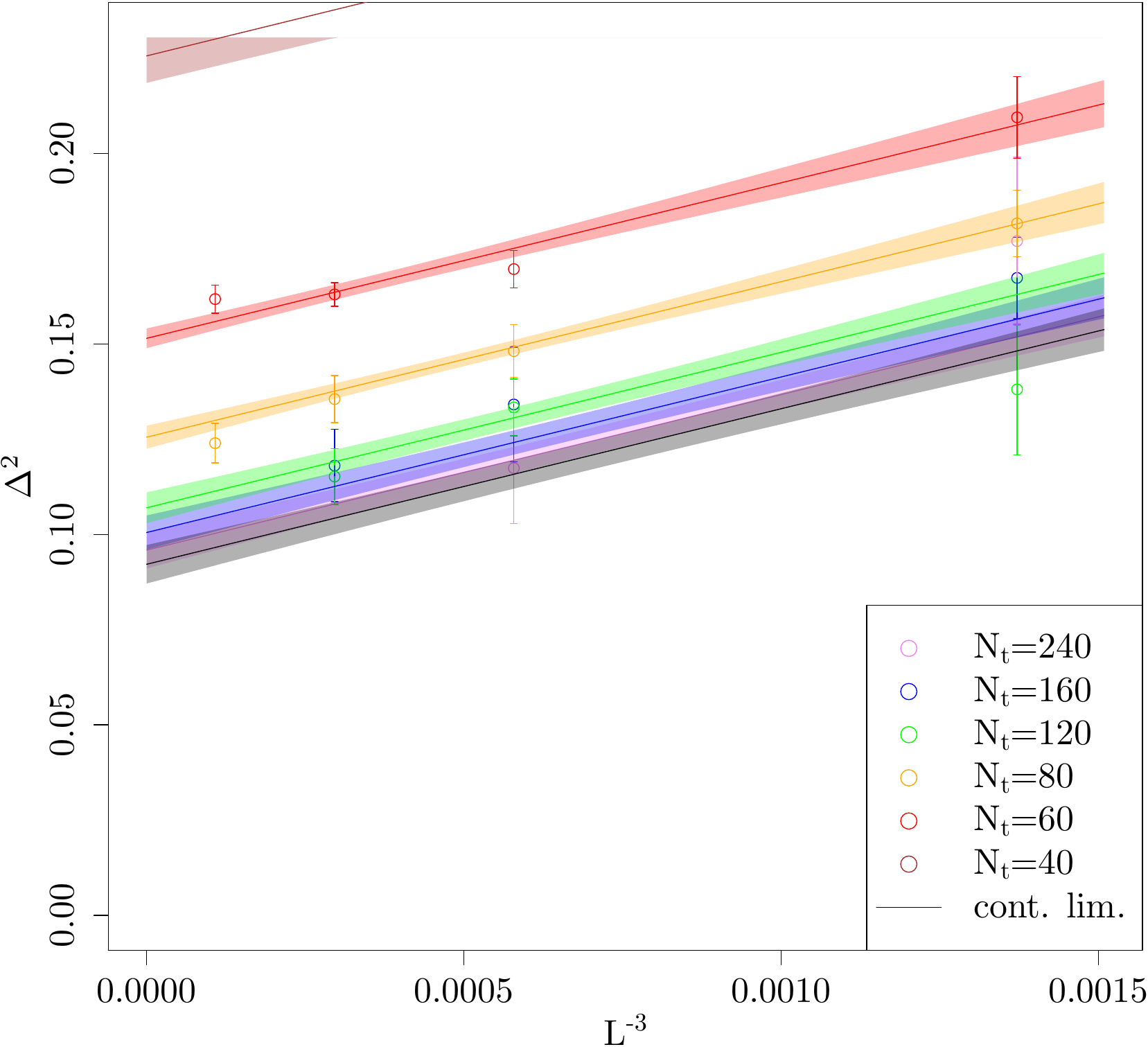}
\includegraphics[width=.45\textwidth]{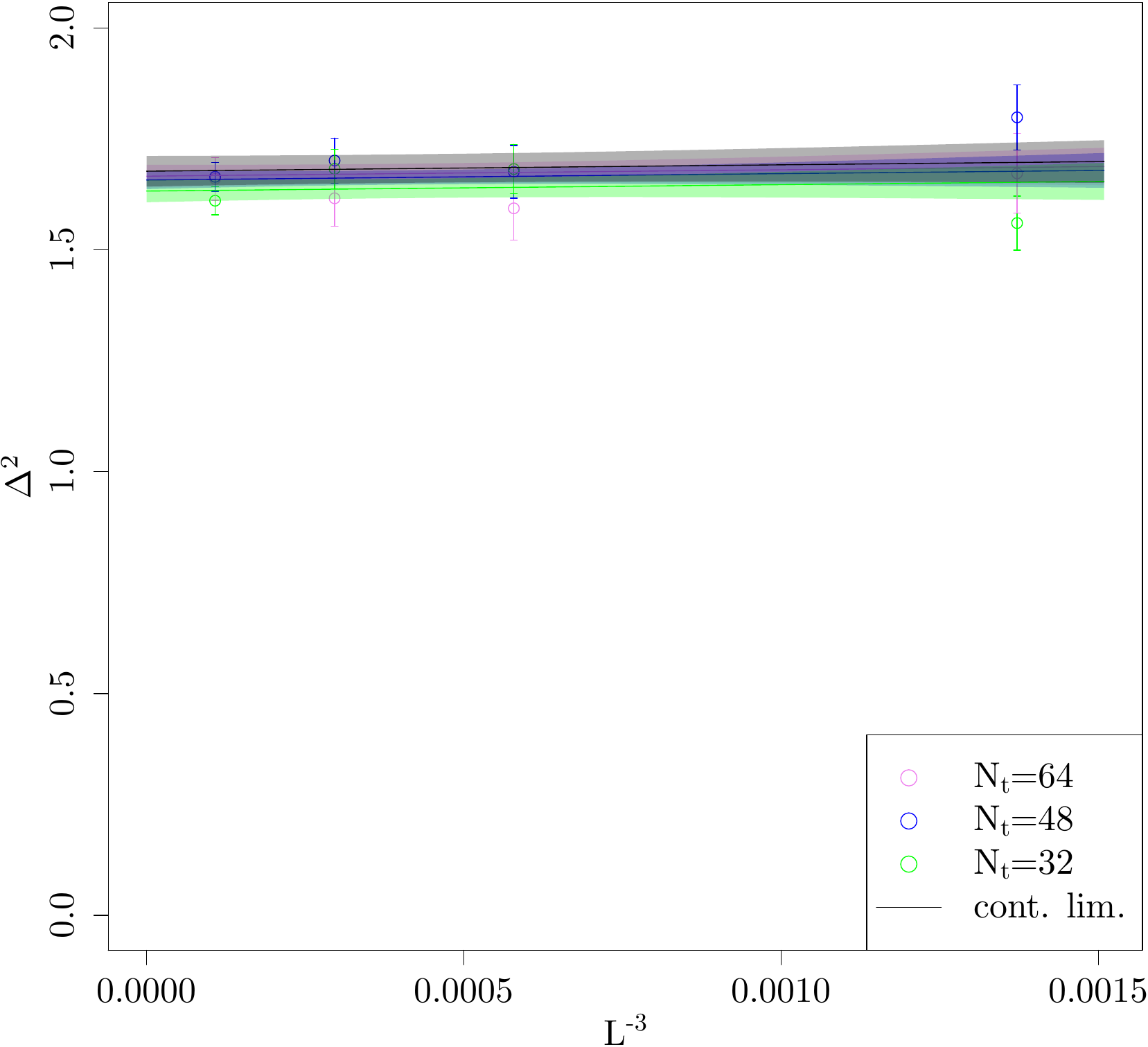}
\caption{Simultaneous two-dimensional fit of $\Delta(N_t,L)$ (in units of $\kappa$) using Eq.~\eqref{eqn_gap_artefacts}, 
for $\kappa\beta=10$ and $U/\kappa=\num{3.9}$ (left panel) and $\kappa\beta=4$ and $U/\kappa=\num{5.0}$ (right panel). Only the extrapolations in $L$ are shown.
Data points for $L<9$ have been omitted from the fits, but not from the plots. 
These fits have $\chi^2/\text{d.o.f.} \simeq \num{0.83}$ and p-value of $\simeq \num{0.62}$ (left panel), and
$\chi^2/\text{d.o.f.} \simeq \num{1.1}$ and p-value of $\simeq \num{0.36}$ (right panel).
\label{fig:finite_size_always_cubic}}
\end{figure}

Higher orders in $U$ influencing the thermal gap are not as easy to calculate, but can in principle be dealt with using
diagrammatic techniques in a finite-temperature Matsubara formalism~\cite{bruus_qft}. As an example, we 
know from \Ref{Giuliani2009} that $v_F^{}$ is influenced (at weak coupling) only at $\ordnung{U^2}$.
Let us finally provide some further numerical evidence for the expected cubic finite-size effects in $L$.
In \Figref{2d_gap_fit}, we have already shown that cubic finite-size effects are a good approximation for $U < U_c$, as expected for
states with small correlation lengths. In \Figref{finite_size_always_cubic}, we show that the cubic behavior in $L$ still holds for $U \simeq U_c$
and $U > U_c$.

\FloatBarrier
\bibliography{cns}

\end{document}